%% file: 2021-CPC.tex
\documentclass{article}

\usepackage{amsmath}
\usepackage{authblk}
\usepackage{color}
\usepackage{graphicx}
\usepackage[utf8]{inputenc}
\usepackage[hidelinks]{hyperref}
\usepackage{algorithm}
\usepackage{algorithmicx}
\usepackage{algpseudocode}
\usepackage{eqparbox}
\usepackage{upgreek}
\usepackage{dsfont}
\usepackage{import}
\usepackage{mathtools}
\usepackage{soul}
\usepackage[normalem]{ulem}

\input{variables.tex}

\newcommand{\Changes}[2][]{#2}

\providecommand{\keywords}[1]{\textbf{\textit{Keywords ---}} #1}

\title
{
  \bf On the implementation of flux limiters in algebraic frameworks
}

\author[1]{Nicol\'as~Valle}
\author[1]{Xavier~\'Alvarez-Farr\'e}
\author[2]{Andrey~Gorobets}
\author[1]{Jes\'us~Castro}
\author[1]{Assensi~Oliva}
\author[1]{F.~Xavier~Trias}

\affil[1]{Heat and Mass Transfer Technological Center, \par Technical University of Catalonia, \par Carrer Colom 11, 08222 Terrassa (Barcelona), Spain}
\affil[2]{Keldysh Institute of Applied Mathematics, \par Russian Academy of Sciences, \par Miusskaya Sq. 4, 125047 Moscow, Russia}

\usepackage[style=numeric, date=year, giveninits=true, maxbibnames=99, doi=true, 
isbn=false, url=false, eprint=false, sorting=none]{biblatex}
\renewbibmacro{in:}{}
\DeclareFieldFormat[article,inproceedings]{doi}{\href{http://doi.org/#1}{\url{http://doi.org/#1}}}
\DeclareFieldFormat[article,inproceedings]{title}{}
\DeclareFieldFormat[article,inproceedings]{journaltitle}{#1}
\DeclareFieldFormat[article,inproceedings]{pages}{#1}
 
\begin{document}
\date{}
\maketitle
\begin{abstract}
\include{abstract}
\end{abstract}
\include{1-Introduction}
\include{2-AlgebraicTopology}
\include{3-FluxLimiter}
\include{4-NumericalResults}
\include{5-Discussion}
\include{6-Conclusions}
\include{acknowledgments}
\printbibliography
\end{document}

%% file: variables.tex
\newcommand{\Reals}{\mathds{R}}             
\newcommand{\set}[1]{{#1}}          
\newcommand{\nel}[1]{|\set{#1}|}            
\newcommand{\msize}[2]{\Reals^{#1 \times #2}}       
\newcommand{\dvec}[1]{\mathbf{#1}}              
\newcommand{\Id}[1]{\mathds{I}_{#1}}        
\newcommand{\Iv}[1]{\mathds{1}_{\set{#1}}}  

\newcommand{\DINC}[1]{\mathsf{E_\set{#1}}}  
\newcommand{\UINC}[1]{\mathsf{B_\set{#1}}}  
\newcommand{\DEG}[1]{\mathsf{W_\set{#1}}}   
\newcommand{\LAPM}[1]{\mathsf{L_\set{#1}}}  
\newcommand{\ADJM}[1]{\mathsf{A_\set{#1}}}  
\newcommand{\UADJM}[1]{\mathsf{A^U_\set{#1}(u)}}
\newcommand{\DADJM}[1]{\mathsf{A^D_\set{#1}}}
\newcommand{\dup}{\dvec{d_U\uptheta}}
\newcommand{\dfc}{\dvec{d_u\uptheta}}
\newcommand{\FLIM}{\mathsf{F(\dvec{r_f})}}  
\newcommand{\UUD}{\mathsf{S_{C \to F}}}
\newcommand{\OUD}{\mathsf{T_{C \to F}}}
\newcommand{\SIGN}{\mathsf{Q(\dvec{u_f})}}  

\newcommand{\topo}[1]{\mathsf{E_\set{#1}}}  
\newcommand{\shift}[1]{\mathsf{\Pi_\set{#1}}}        
\newcommand{\vshift}[1]{\mathsf{\Gamma_\set{#1}}}    
\newcommand{\Nf}{\mathsf{N_\set{F}}}        

\newcommand{\xtc}{\dvec{\uptheta_c}}
\newcommand{\xtcn}{\dvec{\uptheta_c^n}}
\newcommand{\xtcnn}{\dvec{\uptheta_c^{n+1}}}
\newcommand{\xtf}{\dvec{\uptheta_f}}
\newcommand{\xuf}{\dvec{u_f}}
\newcommand{\xrf}{\dvec{r_f}}
\newcommand{\xdf}{\dvec{d_f}}
\newcommand{\xkc}{\dvec{k_c}}
\newcommand{\xnx}{\dvec{n_x}}
\newcommand{\xny}{\dvec{n_y}}
\newcommand{\xnz}{\dvec{n_z}}

\newcommand{\ie}{\textit{i.e.},}
\newcommand{\eg}{\textit{e.g.},}

\newcommand{\oned}{one-dimensional}
\newcommand{\twod}{two-dimensional}
\newcommand{\threed}{three-dimensional}

\newboolean{xbown}
\setboolean{xbown}{true}
\newcommand{\xown}{\ifthenelse{\boolean{xbown}}{\emph{own}\setboolean{xbown}{false}}{own}}
\newboolean{xbouter}
\setboolean{xbouter}{true}
\newcommand{\xouter}{\ifthenelse{\boolean{xbouter}}{\emph{outer}\setboolean{xbouter}{false}}{outer}}
\newboolean{xbinner}
\setboolean{xbinner}{true}
\newcommand{\xinner}{\ifthenelse{\boolean{xbinner}}{\emph{inner}\setboolean{xbinner}{false}}{inner}}
\newboolean{xbinterface}
\setboolean{xbinterface}{true}
\newcommand{\xinterface}{\ifthenelse{\boolean{xbinterface}}{\emph{interface}\setboolean{xbinterface}{false}}{interface}}
\newboolean{xbhalo}
\setboolean{xbhalo}{true}
\newcommand{\xhalo}{\ifthenelse{\boolean{xbhalo}}{\emph{halo}\setboolean{xbhalo}{false}}{halo}}
\newboolean{xbhost}
\setboolean{xbhost}{true}
\newcommand{\xhost}{\ifthenelse{\boolean{xbhost}}{\emph{host}\setboolean{xbhost}{false}}{host}}
\newcommand{\xhosts}{\ifthenelse{\boolean{xbhost}}{\emph{hosts}\setboolean{xbhost}{false}}{hosts}}
\newboolean{xbaccelerator}
\setboolean{xbaccelerator}{true}
\newcommand{\xaccelerator}{\ifthenelse{\boolean{xbaccelerator}}{\emph{accelerator}\setboolean{xbaccelerator}{false}}{accelerator}}
\newcommand{\xaccelerators}{\ifthenelse{\boolean{xbaccelerator}}{\emph{accelerators}\setboolean{xbaccelerator}{false}}{accelerators}}
\newboolean{xbsubvector}
\setboolean{xbsubvector}{true}
\newcommand{\xsubvector}{\ifthenelse{\boolean{xbsubvector}}{\emph{subvector}\setboolean{xbsubvector}{false}}{subvector}}
\newcommand{\xsubvectors}{\ifthenelse{\boolean{xbsubvector}}{\emph{subvectors}\setboolean{xbsubvector}{false}}{subvectors}}
\newboolean{xbsubmatrix}
\setboolean{xbsubmatrix}{true}
\newcommand{\xsubmatrix}{\ifthenelse{\boolean{xbsubmatrix}}{\emph{submatrix}\setboolean{xbsubmatrix}{false}}{submatrix}}
\newcommand{\xsubmatrices}{\ifthenelse{\boolean{xbsubmatrix}}{\emph{submatrices}\setboolean{xbsubmatrix}{false}}{submatrices}}
\newboolean{xbdevice}
\setboolean{xbdevice}{true}
\newcommand{\xdevice}{\ifthenelse{\boolean{xbdevice}}{\emph{device}\setboolean{xbdevice}{false}}{device}}
\newcommand{\xdevices}{\ifthenelse{\boolean{xbdevice}}{\emph{devices}\setboolean{xbdevice}{false}}{devices}}

\newboolean{xbcfd}
\setboolean{xbcfd}{true}
\newcommand{\xcfd}{\ifthenelse{\boolean{xbcfd}}{{computational fluid dynamics (CFD)}\setboolean{xbcfd}{false}}{CFD}}
\newboolean{xbdns}
\setboolean{xbdns}{true}
\newcommand{\xdns}{\ifthenelse{\boolean{xbdns}}{{direct numerical simulation (DNS)}\setboolean{xbdns}{false}}{DNS}}
\newboolean{xbles}
\setboolean{xbles}{true}
\newcommand{\xles}{\ifthenelse{\boolean{xbles}}{{large eddy simulation (LES)}\setboolean{xbles}{false}}{LES}}
\newboolean{xbcfl}
\setboolean{xbcfl}{true}
\newcommand{\xcfl}{\ifthenelse{\boolean{xbcfl}}{{Courant-Friedrichs-Lewy (CFL) condition}\setboolean{xbcfl}{false}}{CFL condition}}
\newboolean{xbisl}
\setboolean{xbisl}{true}
\newcommand{\xisl}{\ifthenelse{\boolean{xbisl}}{{iterative stencil loops (ISL)}\setboolean{xbisl}{false}}{ISL}}
\newboolean{xbslae}
\setboolean{xbslae}{true}
\newcommand{\xslae}{\ifthenelse{\boolean{xbslae}}{{systems of linear algebra equations (SLAE)}\setboolean{xbslae}{false}}{SLAE}}
\newboolean{xbcg}
\setboolean{xbcg}{true}
\newcommand{\xcg}{\ifthenelse{\boolean{xbcg}}{{conjugate gradient (CG)}\setboolean{xbcg}{false}}{CG}}
\newboolean{xbhpc}
\setboolean{xbhpc}{true}
\newcommand{\xhpc}{\ifthenelse{\boolean{xbhpc}}{{high-performance computing (HPC)}\setboolean{xbhpc}{false}}{HPC}}
\newboolean{xbai}
\setboolean{xbai}{true}
\newcommand{\xai}{\ifthenelse{\boolean{xbai}}{{arithmetic intensity (AI)}\setboolean{xbai}{false}}{AI}}
\newboolean{xbflop}
\setboolean{xbflop}{true}
\newcommand{\xflop}{\ifthenelse{\boolean{xbflop}}{{floating-point operations (FLOP)}\setboolean{xbflop}{false}}{FLOP}}
\newcommand{\xflops}{\ifthenelse{\boolean{xbflop}}{{floating-point operations per second (FLOPS)}\setboolean{xbflop}{false}}{FLOPS}}
\newcommand{\xgflop}{\ifthenelse{\boolean{xbflop}}{{giga floating-point operations (GFLOP)}\setboolean{xbflop}{false}}{GFLOP}}
\newcommand{\xgflops}{\ifthenelse{\boolean{xbflop}}{{giga floating-point operations per second (GFLOPS)}\setboolean{xbflop}{false}}{GFLOPS}}
\newcommand{\xtflop}{\ifthenelse{\boolean{xbflop}}{{tera floating-point operations (TFLOP)}\setboolean{xbflop}{false}}{TFLOP}}
\newcommand{\xtflops}{\ifthenelse{\boolean{xbflop}}{{tera floating-point operations per second (TFLOPS)}\setboolean{xbflop}{false}}{TFLOPS}}
\newcommand{\xpflop}{\ifthenelse{\boolean{xbflop}}{{peta floating-point operations (PFLOP)}\setboolean{xbflop}{false}}{PFLOP}}
\newcommand{\xpflops}{\ifthenelse{\boolean{xbflop}}{{peta floating-point operations per second (PFLOPS)}\setboolean{xbflop}{false}}{PFLOPS}}
\newboolean{xbcpu}
\setboolean{xbcpu}{true}
\newcommand{\xcpu}{\ifthenelse{\boolean{xbcpu}}{{central processing unit (CPU)}\setboolean{xbcpu}{false}}{CPU}}
\newcommand{\xcpus}{\ifthenelse{\boolean{xbcpu}}{{central processing units (CPUs)}\setboolean{xbcpu}{false}}{CPUs}}
\newboolean{xbgpu}
\setboolean{xbgpu}{true}
\newcommand{\xgpu}{\ifthenelse{\boolean{xbgpu}}{{graphics processing unit (GPU)}\setboolean{xbgpu}{false}}{GPU}}
\newcommand{\xgpus}{\ifthenelse{\boolean{xbgpu}}{{graphics processing units (GPUs)}\setboolean{xbgpu}{false}}{GPUs}}
\newboolean{xbnuma}
\setboolean{xbnuma}{true}
\newcommand{\xnuma}{\ifthenelse{\boolean{xbnuma}}{{non-uniform memory access (NUMA)}\setboolean{xbnuma}{false}}{NUMA}}
\newboolean{xbdm}
\setboolean{xbdm}{true}
\newcommand{\xdm}{\ifthenelse{\boolean{xbdm}}{{distributed-memory (DM)}\setboolean{xbdm}{false}}{DM}}
\newboolean{xbsm}
\setboolean{xbsm}{true}
\newcommand{\xsm}{\ifthenelse{\boolean{xbsm}}{{shared-memory (SM)}\setboolean{xbsm}{false}}{SM}}
\newboolean{xbmimd}
\setboolean{xbmimd}{true}
\newcommand{\xmimd}{\ifthenelse{\boolean{xbmimd}}{{multiple instruction, multiple data (MIMD)}\setboolean{xbmimd}{false}}{MIMD}}
\newboolean{xbsimd}
\setboolean{xbsimd}{true}
\newcommand{\xsimd}{\ifthenelse{\boolean{xbsimd}}{{simple instruction, multiple data (SIMD)}\setboolean{xbsimd}{false}}{SIMD}}
\newboolean{xbsp}
\setboolean{xbsp}{true}
\newcommand{\xsp}{\ifthenelse{\boolean{xbsp}}{{stream processing (SP)}\setboolean{xbsp}{false}}{SP}}
\newboolean{xbmpi}
\setboolean{xbmpi}{true}
\newcommand{\xmpi}{\ifthenelse{\boolean{xbmpi}}{{message-passing interface (MPI)}\setboolean{xbmpi}{false}}{MPI}}
\newboolean{xbspmv}
\setboolean{xbspmv}{true}
\newcommand{\xspmv}{\ifthenelse{\boolean{xbspmv}}{{sparse matrix-vector product (\texttt{SpMV})}\setboolean{xbspmv}{false}}{\texttt{SpMV}}}
\newboolean{xbaxpy}
\setboolean{xbaxpy}{true}
\newcommand{\xaxpy}{\ifthenelse{\boolean{xbaxpy}}{{linear combination of vectors (\texttt{axpy})}\setboolean{xbaxpy}{false}}{\texttt{axpy}}}
\newboolean{xbkbin}
\setboolean{xbkbin}{true}
\newcommand{\xkbin}{\ifthenelse{\boolean{xbkbin}}{{generalized binary operator (\texttt{kbin})}\setboolean{xbkbin}{false}}{\texttt{kbin}}}
\newboolean{xbdot}
\setboolean{xbdot}{true}
\newcommand{\xdot}{\ifthenelse{\boolean{xbdot}}{{dot product (\texttt{dot})}\setboolean{xbdot}{false}}{\texttt{dot}}}
\newboolean{xbkred}
\setboolean{xbkred}{true}
\newcommand{\xkred}{\ifthenelse{\boolean{xbkred}}{{generalized reduction operator (\texttt{kred})}\setboolean{xbkred}{false}}{\texttt{kred}}}
\newboolean{xbhpcs}
\setboolean{xbhpcs}{true}
\newcommand{\xhpcs}{\ifthenelse{\boolean{xbhpcs}}{{hierarchical parallel code for high-performance computing (HPC\textsuperscript{2})}\setboolean{xbhpcs}{false}}{HPC\textsuperscript{2}}}

%% file: abstract.tex
The use of flux limiters is widespread within the scientific computing community to capture shock discontinuities and are of paramount importance for the temporal integration of high-speed aerodynamics, multiphase flows and hyperbolic equations in general.
Meanwhile, the breakthrough of new computing architectures and the hybridization of supercomputer systems pose a huge portability challenge, particularly for legacy codes, since the computing subroutines that form the algorithms, the so-called kernels, must be adapted to various complex parallel programming paradigms. From this perspective, the development of innovative implementations relying on a minimalist set of kernels simplifies the deployment of scientific computing software on state-of-the-art supercomputers, while it requires the reformulation of algorithms, such as the aforementioned flux limiters.
Equipped with basic algebraic topology and graph theory underlying the classical mesh concept, a new flux limiter formulation is presented based on the adoption of algebraic data structures and kernels. As a result, traditional flux limiters are cast into a stream of only two types of computing kernels: sparse matrix-vector multiplication and generalized pointwise binary operators. The newly proposed formulation eases the deployment of such a numerical technique in massively parallel, potentially hybrid, computing systems and is demonstrated for a canonical advection problem.

%% file: 1-Introduction.tex
\section{Introduction}
\label{sec:Introduction}
The evolution in hardware technologies enables scientific computing to advance incessantly and reach further aims. Nowadays, the use of \xhpc~systems is rather common \Changes[on]{in} the solution of both industrial and academic scale problems. However, many algorithms employed in scientific computing have a very low \xai, which is the ratio of computing work in \xflop~to memory traffic in bytes, hence numerical simulation codes are usually memory-bounded, making processors suffer from serious data starvation \cite{Williams2009, Dongarra2016}. To top it off, the calculations often result in irregular, non-coalescing memory access patterns, reducing the memory access efficiency. Ironically, the memory bandwidth of computing hardware grows much slower than its peak performance, aggravating the problem. All of this motivates the introduction of new parallel architectures with faster and more complex memory configurations into \xhpc~systems.
To take advantage of the increasing variety of computing architectures and the hybridization of \xhpc~systems, the computing subroutines that form the algorithms, the so-called kernels, must be adapted to complex paradigms such as \xdm~and \xsm~\xmimd~parallelism, and \xsp. It also encourages the demand for portable and sustainable implementations of scientific simulation codes \cite{AlFarhan2020}. While portability is an intangible characteristic of software, it may be easy for a developer to have an idea of how difficult it is to rewrite, debug and verify a specific code on its adaptation to a new architecture. On the other hand, sustainability refers to developing reusable and resilient codes. The way a code is conceived at its inception enormously determines the degree to which both properties can be attained.
Traditionally, the development of scientific computing software is based on calculations in \xisl~over a discretized geometry---the mesh. This implementation approach is referred to as \emph{stencil-based}. Despite being intuitive and versatile, the interdependency between algorithms and their computational implementations in stencil applications usually results in a large number of subroutines and introduces an inevitable complexity when it comes to portability and sustainability~\cite{Gysi2015}.
Regarding portability, the complexity of stencil applications motivates the adoption of conservative strategies, which consist of porting (rewriting) the most time-consuming part of an existing code, or even the entire code, to a new architecture but minimizing the structural modifications. In other words, it leads to a partial or complete reimplementation of an existing code. These strategies were common during the rise of general-purpose computing on \xgpus~because they allow for direct comparison studies of both numerical and performance results versus the legacy versions. Well-known commercial \xcfd~codes and open-source platforms offer \xgpu~extensions for solvers of \xslae, which represent a significant part of the overall computing time. This can provide substantial acceleration with compactly localized changes in the code. Such an example can be found in~\cite{Krasnopolsky2016}, where the authors coupled a \xgpu-accelerated library for solving large sparse \xslae~with the OpenFOAM platform and demonstrated performance on up to 128 nodes of a \xgpu-based cluster. The use of \xgpus~in scientific computing is nowadays rather mature, and there are many successful examples in the literature~\cite{Romero2020, Jiang2021, Watanabe2021, Ha2021}.  For instance, the early \xgpu~implementations in \cite{Yamanaka2011}, extended in \cite{Sakane2019}, proved to be two orders of magnitude faster than its \xcpu~counterpart.  Moreover, the solution of two-phase flows on multi-\xgpu~systems \cite{Zaspel2013} was not only faster but also more energy-efficient. An example of a \xgpu~porting of an open-source Navier--Stokes solver, the AFiD code, is found in \cite{Zhu2018}.  Further examples of multi-\xgpu~simulations of supersonic and hypersonic flows can be found in \cite{Bocharov2020}. One of the most impressive \xgpu-based simulations is found in~\cite{Vincent2016}, after \cite{Witherden2014}, on the solution of turbulent flows, reporting a sustained performance of 13.7\xpflops.
Regarding sustainability, the implementation of new physics or numerical methods in a stencil-based framework, or its specialization for different mesh types, usually requires the design of new computing subroutines and data structures. This represents the main drawback of such an approach because the effort is not necessarily accumulative and thus reduces the software's sustainability. To address this, some authors propose domain-specific tools to generalize the stencil computations for specific fields. For instance, a framework that automatically translates stencil functions written in C++ to both \xcpu~and \xgpu~codes is proposed in~\cite{Shimokawabe2016}. However, these generalizations are still heavily restricted by the shape of the stencil they target.
An alternative to stencil implementations is to break the aforementioned interdependency between algorithm and implementation so that the calculations are cast into a minimalist set of kernels. In other words, the idea is to use the classical \xisl~just for building data and leave the calculations to a reduced set of basic operations; thus, legacy codes may be maintained indefinitely as preprocessing tools, and the calculation engines become easy to port and optimize.
By casting discrete operators and mesh functions into sparse matrices and vectors, it has been shown that nearly 90\% of the calculations in a typical \xcfd~algorithm for the \xdns~and \xles~of incompressible turbulent flows boil down to the following basic linear algebra subroutines: \xspmv, \xaxpy~and \xdot~\cite{Oyarzun2017}. Moreover, after the generalizations detailed in Section \ref{sec:AlgebraicImplementation} this value will be raised to 100\%. Hereinafter, we refer to this implementation based on algebraic subroutines as \emph{algebraic} or \emph{algebra-based}. In this algebraic approach, the kernel code shrinks to dozens of lines; the portability becomes natural, and maintaining OpenMP, OpenCL\Changes[]{,} or CUDA implementations takes minor effort. Besides, standard libraries optimized for particular architectures (\eg~cuSPARSE~\cite{NVIDIA2013}, clSPARSE~\cite{Greathouse2016}) can be easily linked in addition to specialised in-house implementations. A similar approach is found in PyFR \cite{Witherden2014}, where the majority of operations are cast in terms of matrix-matrix multiplications linking with appropriate BLAS libraries. In the context of the \xdns, the preconditioned \xcg~method following such an algebraic approach was implemented in~\cite{Oyarzun2014}, and its potential was exploited in~\cite{Borrell2016} to perform petascale \xcfd~simulations.
In our previous work, we proposed a heterogeneous implementation of an algebraic framework~\cite{Alvarez2018}, the \xhpcs, as a portable solution with many potential applications in the fields of computational physics and mathematics. Later, the minimalist design allowed us to easily optimize our framework for \xhpc~systems with \xnuma~configurations~\cite{Alvarez2021}, just by changing the definition of the three kernels and the initialization of vectors and matrices.
While linear schemes fit naturally in the proposed algebraic approach, the implementation of non-linear schemes is not evident. However, as a consequence of Godunov's theorem \cite{Godunov1959,Hirsch1990}, those are required for the treatment of shock discontinuities to avoid unstable discretizations and the onset of wiggles. Such discontinuities are present in many industrial applications and appear in both compressible and multiphase flows, representing a classical problem in numerical analysis since, at least, the 1960's. The construction of stable, second-order (and higher) discretizations, then, requires the adoption of non-linear schemes to exhibit a total variation diminishing (TVD)~\cite{Harten1983} behavior. Among them, flux limiters are a mature and robust method, which has been adopted in a diversity of applications. Sweby \cite{Sweby1984} generalized several limiters and stated the conditions for second-order TVD schemes in a \oned~homogeneous mesh in its well-known \emph{Sweby diagram}. Despite the known inconsistencies that arise when departing from the \oned~homogeneous case \cite{Berger2005, Zeng2016} these techniques have been ported to non-homogeneous Cartesian \cite{Zeng2016} and unstructured \cite{Darwish2003} meshes as well. Advances in this field have also been exploited by the multiphase flow community, particularly for the advection of the marker function \cite{Olsson2005,Balcazar2014b}.
Both the analysis and the implementation of flux limiters are \Changes[tipically]{typically} performed from the aforementioned stencil-based perspective. However, the growing interest of the community in mimetic methods~\cite{Hiemstra2014} unveils an alternative to the implementation of flux limiters. Mimetic methods construct discrete operators directly from the inherent incidence matrices that define the mesh. Adopting such an approach presents an important advantage from both theoretical and practical points of view. On the one hand, this allows for a flawless discrete mimicking of the continuum operators, facilitating the exact conservation of important secondary properties, such as energy \cite{Verstappen2003,Valle2020}, among others. On the other hand, this reduces the implementation to the right combination of a reduced set of algebraic subroutines. Therefore, the present work is devoted to the formulation of flux limiter schemes and their implementation into algebraic frameworks. The proposed implementation will be analyzed from a computational point of view and compared with the classical stencil counterpart, and the benefits of each option will be discussed in depth.
The rest of the paper is organized as follows. In section \ref{sec:Chains} a review of basic concepts of graph theory is briefly summarized in order to provide some context. Section \ref{sec:FluxLimiters} develops a generalization of flux limiters from an algebraic perspective and introduces the matrix-based calculation of the gradient ratio. Section \ref{sec:Results} highlights the capabilities of the method on a well-known \threed~deformation benchmark. Finally, the discussion and conclusions are stated in Sections \ref{sec:Discussion} and \ref{sec:Conclusions}.

%% file: 2-AlgebraicTopology.tex
\section{Algebraic Topology}
\label{sec:Chains}
\begin{figure}[h]
  \centering
  \includegraphics{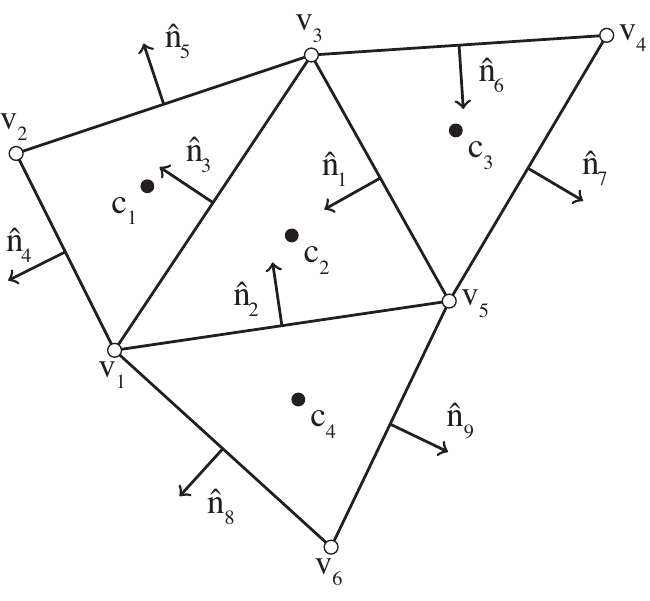}
  \caption{2D mesh composed of cells $c_i$, which are bounded by faces $f_j$ oriented in the direction $\hat{n}_j$. Faces, which collapse to the same entities as edges, are bounded by the set of vertices $v_k$.}
  \label{fig:mesh}
\end{figure}

By using concepts from algebraic topology, mimetic methods preserve the inherent structure of the space, leading to stable and robust discretizations \cite{Robidoux2011,Hiemstra2014}. However, the development of such techniques is out of the scope of this paper, where we rather focus on exploiting the relationships between the different entities of the mesh for the construction of flux limiters. The interested reader is referred to \cite{Lipnikov2014} and references therein.
Given whatever space of interest $\Omega$, we can equip it with a partition of unity, namely a mesh $\set{M}$, by bounding the group of cells, $\set{C}$, with faces, $\set{F}$; those with the set of edges, $\set{E}$, and finally those with the set of vertices, $\set{V}$. In this sequence, groups are related to the next element of the sequence by means of the boundary operator $\partial$. This is know as a chain complex \cite{Robidoux2011,Tonti2014}. A \twod~example can be seen in Figure \ref{fig:mesh}, where faces and edges collapse into the same entity.
In every space $\set{Q}$, discrete variables are arranged in arrays such as $\dvec{\uptheta_q} \in \Reals^{\nel{Q}}$. The relationship between the bounding elements of a geometric entity can be cast in oriented incidence matrices, $\DINC{E \to V}$, $\DINC{F \to E}$ and $\DINC{C \to F}$, corresponding to edge-to-vertex, face-to-edge and cell-to-face, respectively. The corresponding incidence matrices for the mesh depicted in Figure \ref{fig:mesh} read:
\begin{equation}
  \DINC{F \to V} =
  \begin{pmatrix}
     0 & +1 & +1 & +1 &  0 &  0 &  0 & -1 &  0 \cr  
     0 &  0 &  0 & -1 & +1 &  0 &  0 &  0 &  0 \cr
    -1 &  0 & -1 &  0 & -1 & -1 &  0 &  0 &  0 \cr  
     0 &  0 &  0 &  0 &  0 & +1 & -1 &  0 &  0 \cr
    +1 & -1 &  0 &  0 &  0 &  0 & -1 &  0 & +1 \cr  
     0 &  0 &  0 &  0 &  0 &  0 &  0 & +1 & -1 \cr  
  \end{pmatrix},
  \label{eqn:Tvf}
\end{equation}
\begin{equation}
  \DINC{C \to F} =
  \begin{pmatrix}
     0 & -1 & +1 &  0 \cr
     0 & -1 &  0 & +1 \cr
    -1 & +1 &  0 &  0 \cr
    +1 &  0 &  0 &  0 \cr
    +1 &  0 &  0 &  0 \cr
     0 &  0 & -1 &  0 \cr
     0 &  0 & +1 &  0 \cr
     0 &  0 &  0 & +1 \cr
     0 &  0 &  0 & +1 \cr
  \end{pmatrix}.
  \label{eqn:Tfc}
\end{equation}
Conversely, we can define $\DINC{F \to C}$, $\DINC{E \to F}$ and $\DINC{V \to E}$, for the face-to-cell, edge-to-face and vertex-to-edge incidence matrix. Such converse incidence matrices are obtained by transposition:
\begin{align*}
  \DINC{F \to C} &= \DINC{C \to F}^T, \\
  \DINC{E \to F} &= \DINC{F \to E}^T, \\
  \DINC{V \to E} &= \DINC{E \to V}^T.
\end{align*}
Incidence matrices represent the boundary operator between one element of the chain and the next one. Following the example of Figure \ref{fig:mesh}, $\DINC{F \to C}$ and $\DINC{V \to F}$ provide with the orientation of the boundary faces $f_j$ for every cell $c_i$ and the boundary vertices $v_k$ for every face $f_j$. Incidence matrices also play an essential role in preserving properties of the discrete space. In particular, they form an exact sequence. Exact sequences are those such that the application of the boundary operator twice results in 0. This can be verified by checking $\DINC{F \to V}\DINC{C \to F} = 0$. This property is shared by its continuum counterpart, the de Rahm cohomology \cite{Hiemstra2014}, which is the ultimate responsible of the following vector calculus identities \cite{Robidoux2011}:
\begin{align}
  \nabla \times \nabla       & \equiv 0, \\
  \nabla \cdot \nabla \times & \equiv 0.
  \label{eqn:VecCalcId}
\end{align}
These are powerful identities that mimetic methods preserve by construction. For an extended review of the relationship between the continuum and the discrete counterparts, the reader is referred to \cite{Robidoux2011, Hiemstra2014} and references therein.
In addition to provide a suitable platform for the construction of appropriate mimetic methods, the relations contained in incidence matrices can be studied from a graph theory perspective. 
A straightforward use of incidence matrices allows to compute differences across faces. The fact that differences lie in a different space (faces) than variables (cells) is an inherent property of such an approach:
\begin{equation}
  \Delta \xtc = \DINC{C \to F} \xtc.
  \label{eqn:delta}
\end{equation}
Particularly useful is the construction of undirected incidence matrices ($\UINC{Q \to S}$), which are built by taking the absolute value of the elements of the directed ones ($\DINC{Q \to S}$). Considering the index notation between a generic space $\set{Q}$ (\eg~cells, faces) and its boundary $\set{S}$ (\eg~faces, edges), we could proceed as follows:
\begin{equation}
  [\UINC{Q \to S}]_{sq} = b_{sq} = \lvert e_{sq} \rvert,
  \label{eqn:UnOrInMatrix}
\end{equation}
where $e_{sq} = [\DINC{Q \to S}]_{sq}$.
Similarly, one can proceed to compute the degree matrix of the graph, which accounts for the number of connections that an entity has (\eg~the number of cells in contact with a face). Degree matrices are always diagonal and the value of the diagonal elements is obtained as follows:
\begin{equation}
  \DEG{QQ} = diag(\UINC{S \to Q}\Iv{S}).
  \label{eqn:DegMatrix}
\end{equation}
In particular, undirected incidence matrices can be used to construct suitable shift operators \cite{Trias2014}:
\begin{equation}
  \shift{C \to F} = \DEG{FF}^{-1} \UINC{C \to F}.
  \label{eqn:sCF}
\end{equation}
This provides with a simple face-centered interpolation, weighted with the number of adjacent faces. Note that by taking this approach, boundaries are inherently included from the graph information.
The use of such $\shift{C \to F}$ is restricted to scalar fields. However, following \cite{Trias2014}, this can be readily extended to vector fields as follows. First, the discretization of a continuum vector field, $\vec{u} = (u_1,\dots,u_d)^T$, is arranged in a single column vector, $\dvec{u_c} \in \Reals^{d\nel{C} \times 1} = (\dvec{u_1}, \dots, \dvec{u_d})^T$, where $\dvec{u_i} = ((u_i)_1, (u_i)_2, ..., (u_i)_{\nel{C}})^T$ are the vectors containing the components corresponding to the $i$th spatial direction. Note that $d$ is the number of spatial dimensions of the problem. Next, the interpolator can be extended component-wise by applying the Kronecker product with the identity matrix of size $\msize{d}{d}$. The final ensemble is as follows:
\begin{equation}
  \vshift{C \to F} = \Id{d} \otimes \shift{C \to F}.
  \label{eqn:vCF}
\end{equation}
Similarly, normal vectors can be cast into a $\msize{\nel{F}}{d\nel{F}}$ matrix by arranging $d$ diagonal matrices, corresponding to every component of the face vector, next to each other as $\Nf = (N_1 | \dots | N_d)$ \cite{Trias2014}. The \twod~$\Nf$ matrix corresponding to the mesh depicted in Figure \ref{fig:mesh} reads:
\begin{equation}
  \Nf =
  \begin{pmatrix}
    n_{1x} & 0      & \dots  & 0      & n_{1y} & 0      & \dots  & 0      \cr
    0      & n_{2x} & \dots  & 0      & 0      & n_{2y} & \dots  & \vdots \cr
    \vdots & 0      & \ddots & \vdots & \vdots & 0      & \ddots & \vdots \cr
    0      & 0      & \dots  & n_{9x} & 0      & 0      & \dots  & n_{9y} \cr
  \end{pmatrix}.
  \label{eqn:normal}
\end{equation}
In such a way, it is straightforward to either project a discrete vector as $\Nf\dvec{x_f}$, or to vectorize a scalar quantity as  $\Nf^T\dvec{s_f}$, provided that both are stored at the faces. An accurate discussion about the construction of this matrix can be found in \cite{Trias2014}.
Other basic matrices derived from the graph are the graph Laplacian ($\LAPM{CC}$) and the adjacency matrix ($\ADJM{CC}$):
\begin{align}
  \label{eqn:LaplacianMatrix}
  \LAPM{CC} & = \UINC{F \to C}\UINC{C \to F},\\
  \label{eqn:AdjacencyMatrix}
  \ADJM{CC} & = \DEG{CC} - \LAPM{CC}.
\end{align}
Both are constructed based on the incidence matrices and provide information about the propagation of information along the graph. They are constructed by connecting cells to its neighbors through its bounding faces.
In summary, the constructor of such operators provides with tools able to relate different elements of the graph between each others.  Equipped with such basic concepts, the development of higher level operators can proceed as in the following section.

%% file: 3-FluxLimiter.tex
\section{Flux Limiters}
\label{sec:FluxLimiters}
\begin{figure}[h]
  \centering
  \includegraphics[width=0.4\textwidth]{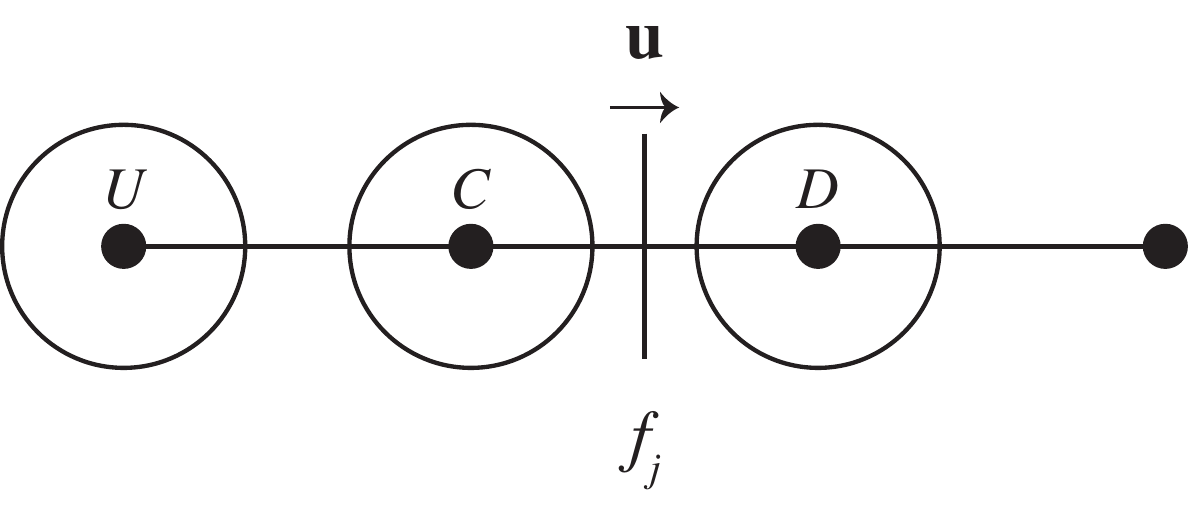}
  \caption{Classical stencil for the computation of the gradient ratio at face $f_j$. $U$, $C$ and $D$ correspond to the upstream, centered and downstream nodes.}
  \label{fig:FL-classical}
\end{figure}

The solution of hyperbolic problems in finite volume methods when sharp discontinuities are present requires the use of high resolution schemes in order to attain second order approximations. In turn, the construction of such schemes is reduced to the appropriate reconstruction of the flux at the faces. Prone to introduce numerical instabilities, such a reconstruction requires an appropriate flux reconstruction strategy in order to guarantee TVD behavior (\ie~such that no new minima or maxima are introduced). This is attained by limiting the flux at cell's boundaries by means of a flux limiter function.
Typically, flux limited schemes are stated in the following form \cite{Sweby1984}:
\begin{eqnarray}
  \theta_f = \theta_C + \Psi(r)\left( \frac{\theta_D-\theta_C}{2} \right),
  \label{eqn:FL-UP}
\end{eqnarray}
where $\theta_C$ and $\theta_D$ stand for the centered and downwind values of $\theta$ according to the velocity field $u$ and $\Psi(r)$ is the flux limiter function. Figure \ref{fig:FL-classical} depicts this situation.
From a physical point of view, this is equivalent to the introduction of some sort of artificial diffusion which stabilizes the method at the expense of smearing out its profile. This can be easily seen by rewriting the classical stencil-based formulation stated in equation (\ref{eqn:FL-UP}) into:
\begin{eqnarray}
  \theta_f = \frac{\theta_C + \theta_D}{2} + \frac{\Psi(r)-1}{2} \left( \theta_D - \theta_C \right),
  \label{eqn:FL-CD}
\end{eqnarray}
where $(\Psi(r)-1)/2$ stands for the artificial diffusion added to a classical symmetry-preserving scheme.
The limiting approach has been used by several authors \cite{Hirsch1988,Hirsch1990} who over the years developed several discontinuity sensors in order to limit the dissipation to the region near the shock. Among all discontinuity sensors, the most popular is the use of the gradient ratio. Following the nomenclature in Figure~\ref{fig:FL-classical}, this is defined as follows~\cite{Sweby1984}:
\begin{equation}
  r_f = \frac{\Delta_U\theta}{\Delta_u\theta}
  = \frac{\theta_C - \theta_U}{\theta_D - \theta_C},
  \label{eqn:gradientRatio}
\end{equation}
where $\Delta_U\theta$ is the gradient of $\theta$ at the upwind face while $\Delta_u\theta$ correspond with the gradient at the face of interest. Both differences are taken as positive in the flow direction, defined by the sign of the velocity field $\xuf$.
This provides with an intuitive description of where discontinuities are and its order of magnitude, at the time that keeps a compact stencil. In addition, it allows to, after proper manipulation by the flux limiter itself, limit the flux in a way which can be interpreted as a diffusion-like term. This is known as ``upwinding'' as it has the same effect as recovering the 1st order upwind discretization near shocks.
TVD conditions in terms of the gradient ratio where stated by Harten in 1983 \cite{Harten1983} for \oned~homogeneous grids. These conditions where used by Sweby in 1984 \cite{Sweby1984} to state 2nd order \emph{and} TVD conditions for different forms of flux limiters. This idea has been extended, with several degrees of accuracy, to multidimensional and irregular grids \cite{Berger2005, Zeng2016}, among others.
%
\subsection{Algebraic Formulation}
\label{sec:AlgebraicFormulation}
As stated previously, the discretization of equation (\ref{eqn:FL-CD}) may benefit from the adoption of an algebraic approach. In this regard, it can be easily extended to the whole computational domain as:
\begin{eqnarray}
  \xtf = \left(  \shift{C \to F} + \FLIM \SIGN \topo{C \to F}\right) \xtc,
  \label{eqn:FL-operator}
\end{eqnarray}
where $\xtf \in \Reals^{\nel{F}}$ and $\xtc \in \Reals^{\nel{C}}$ are the vectors holding all the values of $\theta_f$ and $\theta_c$, $\shift{C \to F} \in \msize{\nel{C}}{\nel{F}}$ is the cell-to-face interpolation defined in equation (\ref{eqn:sCF}), $\FLIM \in \msize{\nel{C}}{\nel{F}}$ is the diagonal matrix absorbing the artificial diffusion introduced in equation (\ref{eqn:FL-CD}), $\SIGN \in \msize{\nel{F}}{\nel{F}}$ is the diagonal matrix taking the proper sign of the velocity at the faces and $\topo{C \to F}$ takes the difference across them as in equation (\ref{eqn:delta}).
At this point, we may be tempted to analyze the construction of new flux limiters by means of basic algebra concepts. In particular, to bound its spectrum by means of Gershgorin's theorem or to check its entropy conditions \cite{Osher1984}, among others. The interested reader is referred to B\'aez et al.~\cite{Baez2016}, where a similar approach is taken for spatial filters.
While both $\shift{C \to F}$ and $\topo{C \to F}$ are readily available from the background stated in section \ref{sec:Chains}, the construction of $\FLIM$ by means of basic algebraic operations solely is addressed next.
Because flux limiter functions $\FLIM$ depend only on the local value of $\xrf$ and we defer the details on the implementation of the pointwise operations to section \ref{sec:AlgebraicImplementation}, the problem is turned into the accurate computation of $\xrf$ at faces. There has been several approaches \cite{Darwish2003,Berger2005} to the construction of $\xrf$ in terms of a least-squares reconstructed gradient. However, the implementation of such schemes can be cumbersome and may not, eventually, recover the \oned~homogeneous solution when a homogeneous structured mesh is used.
The construction of the gradient ratio will proceed first by the separate calculation of both the numerator ($\Delta_U\theta$) and the denominator ($\Delta_u\theta$) of equation (\ref{eqn:gradientRatio}), then computed as:
\begin{equation}
  [\xrf]_i = \frac{[\dup]_i}{[\dfc]_i},
  \label{eqn:rf}
\end{equation}
where $\dfc \in \Reals^{\nel{F}}$ is the face-centered vector holding the difference across the face taken in the direction stated by $\xuf$, while $\dup \in \Reals^{\nel{F}}$ holds the upstream differences according, again, to $\xuf$.
In this approach we propose to employ symmetry-preserving gradients (see \cite{Verstappen2003}) into the calculation of both face-centered and upstream gradients in order to preserve, as much as possible, the mimetic properties of the approach. In addition, we aim at recovering the Cartesian formulation as in \cite{Balcazar2014b}.
Before any calculation, the sign matrix ($\SIGN$) is constructed by assigning to a $\msize{\nel{F}}{\nel{F}}$ diagonal matrix +1 for a positive velocity and -1 for a negative one:
\begin{equation}
  [diag\left(\SIGN\right)]_i = sign([\xuf]_i).
  \label{eqn:sign}
\end{equation}
This allows a straightforward calculation of the velocity-oriented gradient at the face as follows:
\begin{equation}
  \dfc = \SIGN \DINC{C \to F}\xtc,
  \label{eqn:Delta_face}
\end{equation}
where $\SIGN$ is used to provide the right direction in which the difference is taken according to the velocity field.
The construction of the upwind difference $\dup$ is more involved. The idea is to construct a partial adjacency matrix which only considers upstream faces, namely the upstream adjacency matrix, $\UADJM{FF}$, which is responsible to garner upstream information and will be defined further in this paper.
\begin{figure}[h]
  \centering
  \includegraphics{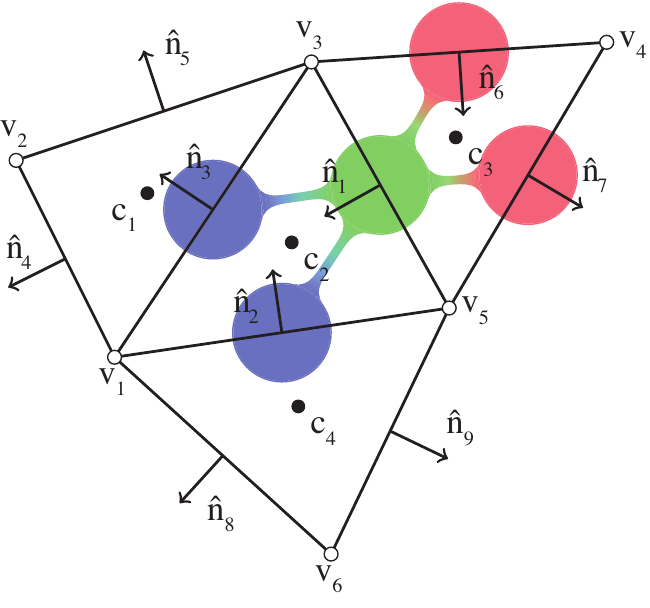}
  \caption{Upstream (red) and downstream (blue) adjacent faces for face $f_1$ with respect to the positive component of velocity at face $f_1$. The selection of the right ones in $\UADJM{FF}$ will ultimately depend on 
  $\SIGN$.}
  \label{fig:AdjacentFaces}
\end{figure}
We proceed as follows: $\topo{C \to F}$ is used to compute the difference across every face according to equation (\ref{eqn:delta}). In order to assess the contribution of every neighboring face to the face of interest, differences are vectorized with its corresponding normal using $\Nf^T$ and added all together with $\UADJM{FF}$. Finally, the resulting value is projected over the normal of the face of interest by means of $\Nf$. The overall construction of the operator is then:
\begin{equation}
  \dup = \Nf \left ( \Id{d} \otimes \UADJM{FF} \right ) \Nf^T \topo{C \to F} \xtc,
  \label{eqn:Delta_Upwind}
\end{equation}
where, similarly to equation (\ref{eqn:vCF}), we reuse the $\UADJM{FF}$ operator for all spatial dimensions. Face normal matrices $\Nf$, defined as in equation (\ref{eqn:normal}), are used for both vectorization and projection of the neighboring differences. In this way, orthogonal meshes recover the original \oned~formulation, whereas unstructured ones are handled inherently by the incidence matrix.
We are now left with construction of the upstream adjacency matrix, $\UADJM{FF}$, which may look cumbersome at a first glance. It can be assembled from other simpler matrices:
\begin{equation}
  \UADJM{FF} = \frac{1}{2} \left ( \ADJM{FF} - \SIGN \DADJM{FF} \right ),
  \label{eqn:UpstreamAdjacency}
\end{equation}
where $\ADJM{FF}$ is the face adjacency matrix, $\DADJM{FF}$ is a ``directed adjacency matrix'', which will be introduced below, and $\SIGN$ is the already defined velocity sign matrix. The strategy for the construction of $\UADJM{FF}$ is to add the contribution of all neighboring faces irrespective of the flow direction and then use $\SIGN \DADJM{FF}$, to remove the downwind ones depending on the values of $\SIGN$. Note that both $\ADJM{FF}$ and $\DADJM{FF}$ are constant matrices and that the only matrix that needs to be updated according to $\xuf$ is $\SIGN$.
The construction of $\ADJM{FF}$ proceeds similarly to equation (\ref{eqn:AdjacencyMatrix}) as follows:
\begin{equation}
  \ADJM{FF} = \DEG{FF} - \UINC{C \to F}\UINC{F \to C}.
  \label{eqn:AdjacencyFaces}
\end{equation}
As seen in section \ref{sec:Chains}, the adjacency matrix is symmetric and contains non-negative entries only. Following on the example depicted in Figure \ref{fig:AdjacentFaces}, the corresponding $\ADJM{FF}$ reads:
\begin{equation}
  \ADJM{FF} =
  \begin{pmatrix}
     0 & +1 & +1 &  0 &  0 & +1 & +1 &  0 &  0 \cr
    +1 &  0 & +1 &  0 &  0 &  0 &  0 & +1 & +1 \cr
    +1 & +1 &  0 & +1 & +1 &  0 &  0 &  0 &  0 \cr  
     0 &  0 & +1 &  0 & +1 &  0 &  0 &  0 &  0 \cr  
     0 &  0 & +1 & +1 &  0 &  0 &  0 &  0 &  0 \cr
    +1 &  0 &  0 &  0 &  0 &  0 & +1 &  0 &  0 \cr
    +1 &  0 &  0 &  0 &  0 & +1 &  0 &  0 &  0 \cr
     0 & +1 &  0 &  0 &  0 &  0 &  0 &  0 & +1 \cr  
     0 & +1 &  0 &  0 &  0 &  0 &  0 & +1 &  0 \cr
  \end{pmatrix}.
  \label{eqn:ADJM}
\end{equation}
On the other hand, the construction of $\DADJM{FF}$ allows us to distinguish neighboring faces which lie, according to the face normal, behind or ahead of the face in question. This requires the inclusion of the directed incidence matrix $\DINC{C \to F}$ into the calculation of the adjacency matrix as:
\begin{equation}
  \DADJM{FF} = \DINC{C \to F}\UINC{F \to C},
  \label{eqn:DirectedAdjacencyFaces}
\end{equation}
which provides with the following matrix:
\begin{equation}
  \DADJM{FF} =
  \begin{pmatrix}
     0 & +1 & +1 &  0 &  0 & -1 & -1 &  0 &  0 \cr
    +1 &  0 & +1 &  0 &  0 &  0 &  0 & -1 & -1 \cr
    -1 & -1 &  0 & +1 & +1 &  0 &  0 &  0 &  0 \cr  
     0 &  0 & -1 &  0 & -1 &  0 &  0 &  0 &  0 \cr  
     0 &  0 & -1 & -1 &  0 &  0 &  0 &  0 &  0 \cr
    +1 &  0 &  0 &  0 &  0 &  0 & +1 &  0 &  0 \cr
    -1 &  0 &  0 &  0 &  0 & -1 &  0 &  0 &  0 \cr
     0 & -1 &  0 &  0 &  0 &  0 &  0 &  0 & -1 \cr  
     0 & -1 &  0 &  0 &  0 &  0 &  0 & -1 &  0 \cr
  \end{pmatrix}.
  \label{eqn:DADJM}
\end{equation}
Note that $\DADJM{FF}$ has the same pattern as $\ADJM{FF}$ but entries corresponding to faces located upstream (with respect to the face normal direction) contain $-1$ whereas those located downstream contain $+1$, as shown in Figure \ref{fig:AdjacentFaces}.
However, the choice of upstream/downstream faces should depend on the faces local velocity and not on its arbitrary choice of face normal. The product $\SIGN \DADJM{FF}$, corrects this by inverting the sign of the rows corresponding to the faces whose velocity component is not aligned with the face normal. The result is a correct choice of upstream and downstream faces according to the local face velocity.
Finally, the combination of $\DADJM{FF}$ and $\ADJM{FF}$ in equation (\ref{eqn:Delta_Upwind}) results in:
\begin{equation}
  \dup  = \Nf \left ( \Id{d} \otimes \frac{1}{2} \left ( \SIGN \DADJM{FF} - \ADJM{FF} \right ) \right ) \Nf^T \topo{C \to F} \xtc.
  \label{eqn:FinalDeltaUpstream}
\end{equation}
While equation (\ref{eqn:FinalDeltaUpstream}) succeeds at selecting the proper upstream faces, its direct implementation involves many redundant operations that may result in an unnecessary overhead. For this reason, the computation of $\dup$ is rearranged as follows:
\begin{equation}
  \dup  = \left( \SIGN \UUD + \OUD  \right)\xtc,
  \label{eqn:SimplifiedDeltaUpstream}
\end{equation}
where we introduce the new matrices $\UUD = \frac{1}{2} \Nf \left ( \Id{d} \otimes \DADJM{FF} \right ) \Nf^T \topo{C \to F}$ and $\OUD = \frac{1}{2} \Nf \left ( \Id{d} \otimes \ADJM{FF} \right ) \Nf^T \topo{C \to F}$, which can be precomputed at the beginning of the simulation.
%
\subsection{Algebraic Implementation}
\label{sec:AlgebraicImplementation}
In previous works of Oyarzun et al.~\cite{Oyarzun2017} and \'Alvarez et al.~\cite{Alvarez2018}, an algebra-based implementation model was proposed for the \xdns~and \xles~of incompressible turbulent flows such that the algorithm of the time-integration phase reduces to a set of only three algebraic kernels: \xspmv, \xaxpy~and \xdot. However, a close look at Equations \ref{eqn:rf} and \ref{eqn:sign}, for instance, reveals that this set is insufficient to fulfill the implementation of the flux limiter because it comprises non-linear operations.
Nevertheless, instead of being an inconvenience, this encourages us to demonstrate the high potential of our algebraic strategy again. We propose the generalization of the \xaxpy~via the introduction of a \xkbin~that performs any given pointwise arithmetic calculation such that:
\begin{equation}
  y_i \leftarrow y_i \circ f(x_i).
  \label{eqn:BINARY-OPERATOR}
\end{equation}
This binary operator can easily map to the required kernels by defining $\circ$ and $f(x_i)$ as outlined in Table \ref{tab:BINARY-KERNELS}. Similarly, the \xdot~kernel can be turned into a \xkred,
\begin{equation}
  r \leftarrow r \circ f(x_i),
  \label{eqn:REDUCTION-OPERATOR}
\end{equation}
which can easily represent any required reduction operation such as the calculation of the norm of a vector or the \xcfl. However, the details of the \xkred~are out of the scope of this work because it is not required for the implementation of the flux limiter.
\begin{table}[ht!]
  \caption{Particularizations of the operator $\circ$ and pointwise function $f(x_i)$ to represent various kernels using the generalized binary operator described in Equation \ref{eqn:BINARY-OPERATOR}. The \texttt{superbeexty} corresponds to the SUPERBEE flux limiter \cite{Sweby1984}.}
  \begin{center}
    \begin{tabular}{c c c r}
      $\circ$  & $f(x_i)$ & AI                & Resulting Kernel \\ \hline
      $+$      & $ax_i$   & $1/12$            & \texttt{axpy}    \\
      $\times$ & $ax_i$   & $1/12$            & \texttt{axty}    \\
      $/$      & $ax_i$   & $1/12$            & \texttt{axdy}    \\
      $\times$ & $x_i > 0 ? +1 : -1$ & $1/12$ & \texttt{signxty} \\
      $\times$ & $0.5(max(0, max(min(1, 2x_i), min(x_i, 2)))-1)$ & $7/24$ & \texttt{superbeexty}\\ \hline
    \end{tabular}
  \end{center}
  \label{tab:BINARY-KERNELS}
\end{table}
From a computational point of view, this kernel generalization does not alter the implementation of the original \xaxpy: it still performs simple, pointwise arithmetic operations over the vector elements and provides uniform, aligned and coalescing memory accesses which suits the \xsimd~and \xsp~paradigms perfectly. Therefore, having already efficient implementations of \xaxpy~for different architectures, the implementation of \xkbin~is straightforward (\eg~consider the use of function pointers, templates, macros, among others).
On the other hand, the arithmetic intensity of this new kernel is not a fixed value anymore, as shown in Table \ref{tab:BINARY-KERNELS}. While the \xai~of the \xaxpy~was $1/12$ \xflop~per byte (one product and one addition per three 8-byte values), that of the \xkbin~will depend on the specific arithmetic calculations involved in the function $f(x_i)$. This allows us to significantly increase the \xai~in our calls by means of kernel fusion, reduce the number of intermediate results, and thus reduce the time-to-solution.
The final algorithm for the deployment of a flux limiter in the reconstruction of the variable at faces, $\theta_f$, within our algebra-based framework is described in Algorithm \ref{alg:FLUX-LIMITER-ALG}.
\begin{algorithm}[h]
  \caption{Algorithm for reconstruction of a scalar field at faces, $\xtf$, using the algebraic implementation of a flux limiter.}
  \begin{algorithmic}[1]
    \Require $\xtc$, $\xuf$, $\DINC{C \to F}$, $\UUD$, $\OUD$, $\shift{C \to F}$
    \Ensure $\xtf$
    \State $\dfc \gets \DINC{C \to F} \xtc$
    \Comment{\texttt{SpMV}} \label{alg:FLUX-LIMITER-ALG:line:UPWIND-DIFFERENCE-1}
    \State $\dfc \gets \SIGN \dfc$
    \Comment{\texttt{signxty}} \label{alg:FLUX-LIMITER-ALG:line:UPWIND-DIFFERENCE-2}
    \State $\mathrlap{\xrf}\hphantom{\dfc} \gets \UUD\xtc$
    \Comment{\texttt{SpMV}} \label{alg:FLUX-LIMITER-ALG:line:GRADIENT-RATIO-1}
    \State $\mathrlap{\xrf}\hphantom{\dfc} \gets \SIGN \xrf$
    \Comment{\texttt{signxty}} \label{alg:FLUX-LIMITER-ALG:line:GRADIENT-RATIO-2}
    \State $\mathrlap{\xrf}\hphantom{\dfc} \gets \xrf + \OUD\xtc$
    \Comment{\texttt{SpMV}} \label{alg:FLUX-LIMITER-ALG:line:GRADIENT-RATIO-3}
    \State $\mathrlap{\xrf}\hphantom{\dfc} \gets \xrf/\dfc$
    \Comment{\texttt{axdy}} \label{alg:FLUX-LIMITER-ALG:line:GRADIENT-RATIO-4}
    \State $\mathrlap{\xtf}\hphantom{\dfc} \gets \FLIM \dfc$
    \Comment{\texttt{superbeexty}} \label{alg:FLUX-LIMITER-ALG:line:THETA-AT-FACE-1}
    \State $\mathrlap{\xtf}\hphantom{\dfc} \gets \xtf + \shift{C \to F}\xtc$
    \Comment{\texttt{SpMV}} \label{alg:FLUX-LIMITER-ALG:line:THETA-AT-FACE-2}
  \end{algorithmic}
  \label{alg:FLUX-LIMITER-ALG}
\end{algorithm}
Note that because we are actually interested in the evaluation of $\dup$ rather than in the construction of the operator itself, matrix-matrix products are avoided and successive matrix-vector products are performed. Similarly, we are not interested in the construction of the diagonal matrices $\SIGN$ and $\FLIM$. We will rely on the generalized binary operator instead. Therefore, the evaluation of $\SIGN \dvec{y}$ will be done using the \texttt{signxty} in Table \ref{tab:BINARY-KERNELS} as follows:
\begin{equation}
  [\dvec{y}]_i \leftarrow [\dvec{y}]_i \times ([\xuf]_i > 0 ? +1 : -1).
\end{equation}
Finally, a particular flux limiter function must be chosen on the evaluation of $\FLIM \dvec{y}$. Our framework allows to easily switch between different functions. Hereinafter, we will proceed with the \emph{superbee} \cite{Sweby1984}, represented by \texttt{superbeexty}:
\begin{equation}
  [\dvec{y}]_i \leftarrow [\dvec{y}]_i \times 0.5(max(0, max(min(1, 2[\xrf]_i), min([\xrf]_i, 2)))-1).
\end{equation}
In conclusion, the evaluation of a scalar field at faces, $\xtf$, as in Algorithm \ref{alg:FLUX-LIMITER-ALG} can be fitted in our algebra-based framework by combining two types of computing kernels: \xspmv~and \xkbin.
%
\subsection{Comparison with stencil-based implementations}
\label{sec:ComparisonWithStencil}
\begin{figure}[t]
  \centering
  \includegraphics[width=\textwidth]{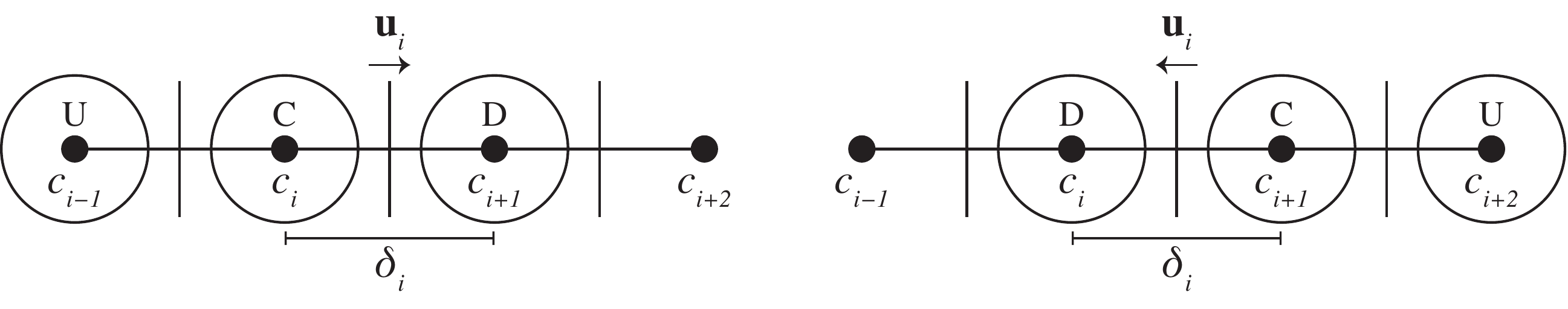}
  \caption{Example of stencils in a \protect\oned~Cartesian grid according to the classical flux limiter approach. The stencil topology varies according to the sign of the velocity field, $\vec{v}$.}
  \label{fig:IMG-DIAGRAM-FLUX-LIMITER-1D}
\end{figure}

Our algebraic implementation for the appropriate reconstruction of the variable\Changes[]{s} at faces, $\xtf$, is now compared with classical, stencil-based approaches. This comparison is conducted \Changes[]{firstly} from a theoretical point of view, assessing the minimum number of \xflop~and memory traffic (in bytes) required in different scenarios. \Changes[]{Note that the actual number of memory accesses during kernel execution depends not only on the algorithm but also on hardware and software features. Therefore, regarding} the memory traffic, two different values are estimated considering the \emph{full-hit} and \emph{full-miss} caseloads\Changes[, that is, an ideal and null temporal locality, respectively]{. The former refers to the best scenario with an ideal temporal locality: multiple accesses to a particular data element are so close in time that its value is always reused from cache. Conversely, the latter considers the worst scenario with a null temporal locality so that every repeated access results in cache-miss and requires a memory load from memory. Thus, these two values result in the interval of effective \xai~of each kernel.}
For the sake of clarity, \Changes[]{in this comparison} we only consider $k$-regular, periodic meshes composed of convex polygons which accomplish the following equality:
\begin{equation}
\nel{F} \approx \frac{k}{2} \nel{C},
\end{equation}
where $\nel{C}$ and $\nel{F}$ represent the number of cells and faces, respectively, and $k$ is the degree of the mesh elements (\ie~the number of neighboring cells). The resulting requirements are listed in Table \ref{tab:FLUX-LIMITER-REQUIREMENTS}, which is described throughout the section.
Let us start from the analysis of the simplest case: the stencil-based calculation of $\theta_f$ in a one-dimensional Cartesian grid, depicted in Figure \ref{fig:IMG-DIAGRAM-FLUX-LIMITER-1D} and described in Algorithm \ref{alg:FLUX-LIMITER-1D}. Every $i$th face is surrounded by two cells, $c_i$ and $c_{i+1}$. For each $i$th face in $\nel{F}$, the sign of the velocity determines the upstream, centered and downstream values of $\theta$, as well as the centered and upstream distances. Following the algorithm, $5\nel{F}$ \xflop~are required for computing the gradient ratio in line \ref{alg:FLUX-LIMITER-1D:line:GRADIENT-RATIO}, $1\nel{F}$ for computing the limiter function in line \ref{alg:FLUX-LIMITER-1D:line:PSI} (this value may vary depending on the limiter function; in this example we have considered the \textit{superbee} limiter \cite{Sweby1984}), and $7\nel{F}$ for computing the value at the face in line \ref{alg:FLUX-LIMITER-1D:line:THETA-AT-FACE}. In the algorithm, three discrete fields are required for the computations: the initial scalar field, $\xtc \in \Reals^{\nel{C}}$, the velocity field, $\xuf \in \Reals^{\nel{F}}$ and the distances between cells, $\xdf \in \Reals^{\nel{F}}$. Besides, the algorithm ensures the calculation of the discrete scalar field at faces, $\xtf \in \Reals^{\nel{F}}$. Thus, considering double-precision values, and given that the number of faces is equal to the number of cells in the one-dimensional case ($k=2 \rightarrow \nel{F} = \nel{C}$), the minimum \xflop~and bytes required are $5\nel{F} + 1\nel{F} + 7\nel{F} = 13\nel{C}$ and $8(\nel{C} + \nel{F} + \nel{F} + \nel{F}) = 32\nel{C}$, respectively. The total memory traffic \Changes[if there was null temporal locality (full-miss)]{in the full-miss caseload} would rise to $48\nel{C}$ because two different values of $\xdf$ and $\xtc$ are \Changes[to be read]{accessed} for every face. Note that this values are slightly reduced in the particular case of uniform meshes: neither the distances array nor its quotient are required, thus omitting $2\nel{F}$ \xflop~in line \ref{alg:FLUX-LIMITER-1D:line:GRADIENT-RATIO} and the access to $\xdf$. On the other hand, the generalization of Algorithm \ref{alg:FLUX-LIMITER-1D} for \threed~Cartesian grids ($k=6 \rightarrow \nel{F} = 3\nel{C}$) is straightforward and rises the computational requirements to $39 \nel{C}$ \xflop~and $80$--$336 \nel{C}$ bytes for non-uniform meshes, or $33\nel{C}$ \xflop~and $56$--$216 \nel{C}$ bytes in the uniform case.
\begin{algorithm}[h]
  \begin{algorithmic}[1]
    \Require $\xtc$, $\xuf$, $\xdf$
      \Comment{8C + 8F + 8F bytes}
    \Ensure $\xtf$
      \Comment{8F bytes}
    \For {$i \gets 1$ to $\nel{F}$}
      \If {$\xuf[i]>0$}
        \State $ \theta_C = \mathrlap{\xtc[i], \quad}\hphantom{\xtc[i+1], \quad{}} \mathrlap{\theta_U}\hphantom{d_U} = \xtc[i-1], \quad \theta_D = \xtc[i+1]$
        \State $ \mathrlap{d_i}\hphantom{\theta_C} = \mathrlap{\xdf[i], \quad}\hphantom{\xtc[i+1], \quad{}} d_U = \xdf[i - 1]$
      \Else
        \State $ \theta_C = \xtc[i+1], \quad \mathrlap{\theta_U}\hphantom{d_U} = \mathrlap{\xtc[i], \quad}\hphantom{\xtc[i-1], \quad{}} \theta_D = \xtc[i+2]$
        \State $ \mathrlap{d_i}\hphantom{\theta_C} = \mathrlap{\xdf[i], \quad}\hphantom{\xtc[i+1], \quad{}} d_U = \xdf[i + 1]$
      \EndIf
      \State $ \mathrlap{r_i}\hphantom{\xtf[i]} \gets (d_i/d_U)(\theta_C - \theta_U)/(\theta_D - \theta_C)$
        \Comment{5F flops} \label{alg:FLUX-LIMITER-1D:line:GRADIENT-RATIO}
      \State $ \mathrlap{\Psi_i}\hphantom{\xtf[i]} \gets limiter(r_f)$
        \Comment{1F flops} \label{alg:FLUX-LIMITER-1D:line:PSI}
      \State $ \xtf[i] \gets (\theta_C + \theta_D)/2 + (\Psi_i - 1)(\theta_D - \theta_C)/2$
        \Comment{7F flops} \label{alg:FLUX-LIMITER-1D:line:THETA-AT-FACE}
    \EndFor
  \end{algorithmic}
  \caption{Stencil-based calculation of $\theta_f$ in a \protect\oned~Cartesian grid.}
  \label{alg:FLUX-LIMITER-1D}
\end{algorithm}
A generalization of the stencil calculation for \Changes[any kind of unstructured mesh]{unstructured meshes} is outlined in Algorithm \ref{alg:FLUX-LIMITER-UNS}. In contrast with the structured algorithm, the indices of neighboring nodes are not predictable\Changes[ in this case and hence the incidence graphs are required]{, so the incidence graphs are required}. For each $i$th face in $\nel{F}$, the sign of the velocity determines the indices of the centred and downstream cells, $c_C$ and $c_D$, according to the cell-to-face incidence graph. Then, for each $j$th face incident to $c_C$ (except $f_i$), its contribution to the upstream gradient, projected over the normal of $f_i$, is accumulated, accounting for $5(k-1)\nel{F}$ and $4(k-1)\nel{F}$ \xflop~in lines \ref{alg:FLUX-LIMITER-UNS:line:PROJECTION} and \ref{alg:FLUX-LIMITER-UNS:line:ACCUMULATION}, respectively. The calculation of $\theta_f$ in lines \ref{alg:FLUX-LIMITER-UNS:line:GRADIENT-RATIO} to \ref{alg:FLUX-LIMITER-UNS:line:THETA-AT-FACE} follows similarly as in Algorithm \ref{alg:FLUX-LIMITER-1D}, and adds $13\nel{F}$ operations. In this case, six discrete fields, one integer list and two incidence graphs are required for the computations. The specific requirements for two k-regular unstructured meshes, $k=2$ (\oned~mesh) and $k=6$ (\threed~hexaedral mesh), are listed in Table \ref{tab:FLUX-LIMITER-REQUIREMENTS}.
\begin{algorithm}[ht!]
  \begin{algorithmic}[1]
    \Require $\xtc$, $\xuf$, $\xdf$, $\xnx$, $\xny$, $\xnz$, 
    \Comment{8C + 40F bytes}
    \par\hskip\algorithmicindent $\xkc$, $\DINC{C \to F}$, $\DINC{F \to C}$
    \Comment{4C + 8F + 8F bytes}
    \Ensure $\xtf$
      \Comment{8F bytes}
    \For {$i \gets 1$ to $\nel{F}$}
      \If {$\xuf[i]>0$}
        \State $c_C = \DINC{C \to F}[i][0]$
        \State $c_D = \DINC{C \to F}[i][1]$
      \Else
        \State $c_C = \DINC{C \to F}[i][1]$
        \State $c_D = \DINC{C \to F}[i][0]$
      \EndIf
      \State $\Delta_U = 0$
      \For {$k \gets 1$ to $\xkc[c_C]$}
        \State $j = \DINC{F \to C}[c_C][k]$
        \If {$j \ne i$}
          \If {$\DINC{C \to F}[j][0] \ne c_C$}
            \State $\mathrlap{c_U}\hphantom{c_U} = \DINC{C \to F}[j][0]$
          \Else
            \State $\mathrlap{c_U}\hphantom{c_U} = \DINC{C \to F}[j][1]$
          \EndIf
          \State \Changes[]{$\mathrlap{d_j}\hphantom{\theta_U} = \xdf[j]$}
          \State $\theta_U = \xtc[c_U]$
          \State $\mathrlap{n_k}\hphantom{\Delta_U} \gets \xnx[i]\xnx[j] + \xny[i]\xny[j] + \xnz[i]\xnz[j]$
          \Comment{5(k-1)F flops} \label{alg:FLUX-LIMITER-UNS:line:PROJECTION}
          \State $\Delta_U \gets \Delta_U + n_k(\theta_C - \theta_U)/d_j$
          \Comment{4(k-1)F flops} \label{alg:FLUX-LIMITER-UNS:line:ACCUMULATION}
        \EndIf
      \EndFor
      \State \Changes[]{$\mathrlap{r_i}\hphantom{\xtf[i]} \gets \Delta_U/((\theta_D - \theta_C)/d_i)$}
      \Comment{3F flops} \label{alg:FLUX-LIMITER-UNS:line:GRADIENT-RATIO}
      \State $\mathrlap{\Psi_i}\hphantom{\xtf[i]} \gets superbee(r_i)$
      \Comment{1F flops} \label{alg:FLUX-LIMITER-UNS:line:PHI}
      \State $\xtf[i] \gets (\theta_C + \theta_D)/2 + (\Psi_i - 1)(\theta_D - \theta_C)/2$
      \Comment{7F flops} \label{alg:FLUX-LIMITER-UNS:line:THETA-AT-FACE}
    \EndFor
  \end{algorithmic}
  \caption{Stencil-based calculation of $\theta_f$ in a generic unstructured grid based on incidence graphs.}
  \label{alg:FLUX-LIMITER-UNS}
\end{algorithm}
Finally, \Changes[]{in} the algebraic implementation \Changes[is]{} outlined in Algorithm \ref{alg:FLUX-LIMITER-ALG}\Changes[. It]{, it} can be observed how \Changes[this algorithm]{it} is completely independent of the mesh type and the numerical method: these characteristics only affect the matrices. The number of calls to \xspmv~and \xkbin~kernels is readily deduced from the algorithm: 4 times each. Four matrices are required for the computations: the differences at faces, $\DINC{C \to F}$, with $2\nel{F}$ non-zero elements, the oriented and unoriented differences, $\UUD$ and $\OUD$, with $2k\nel{F}$ each, and the cell-to-face interpolation, $\shift{C \to F}$, with $2\nel{F}$. In this example, we consider the use of the ELLPACK format \cite{Kincaid1989} in which each non-zero element accounts for 12 bytes (\ie~8 bytes for the coefficient and 4 bytes for the column index). The specific requirements for two k-regular meshes are listed in Table \ref{tab:FLUX-LIMITER-REQUIREMENTS}.
\begin{table}[h]
\small
\caption{Minimum number of \protect\xflop~and memory traffic (in bytes) required per mesh cell for computing the variable at the faces in different scenarios: stencil- and algebra-based implementations on uniform, non-uniform and unstructured \protect\oned~($k=2$) and \protect\threed~($k=6$) meshes.}
\begin{center}
\begin{tabular}{c c c c c c c}
k & Approach     & FLOP  & \multicolumn{2}{c}{Bytes} & \multicolumn{2}{c}{AI} \\ \hline
  &              &       & full-hit     & full-miss  & full-hit & full-miss \\ \hline
2 & Uniform      & $11$  & $32$         & $48$       & $0.344$  & $0.229$   \\ 
2 & Non-uniform  & $13$  & $40$         & $56$       & $0.325$  & $0.232$   \\ 
2 & Unstructured & $20$  & $84$         & $132$      & $0.238$  & $0.152$   \\
2 & Algebraic    & $28$  & $288$        & $352$      & $0.097$  & $0.080$   \\ \hline
6 & Uniform      & $33$  & $80$         & $240$      & $0.413$  & $0.138$   \\ 
6 & Non-uniform  & $39$  & $104$        & $360$      & $0.375$  & $0.108$   \\ 
6 & Unstructured & $168$ & $228$        & $1020$     & $0.737$  & $0.165$   \\
6 & Algebraic    & $180$ & $1408$       & $1984$     & $0.128$  & $0.091$   \\ \hline
\end{tabular}
\end{center}
\label{tab:FLUX-LIMITER-REQUIREMENTS}
\end{table}

%% file: 4-NumericalResults.tex
\section{Numerical \Changes[Results]{Study}}
\label{sec:Results}
\subsection{\Changes[]{Three-dimensional deformation problem}}
\label{sec:Vortex}
\begin{figure}[h]
  \centering
  \includegraphics[width=\textwidth]{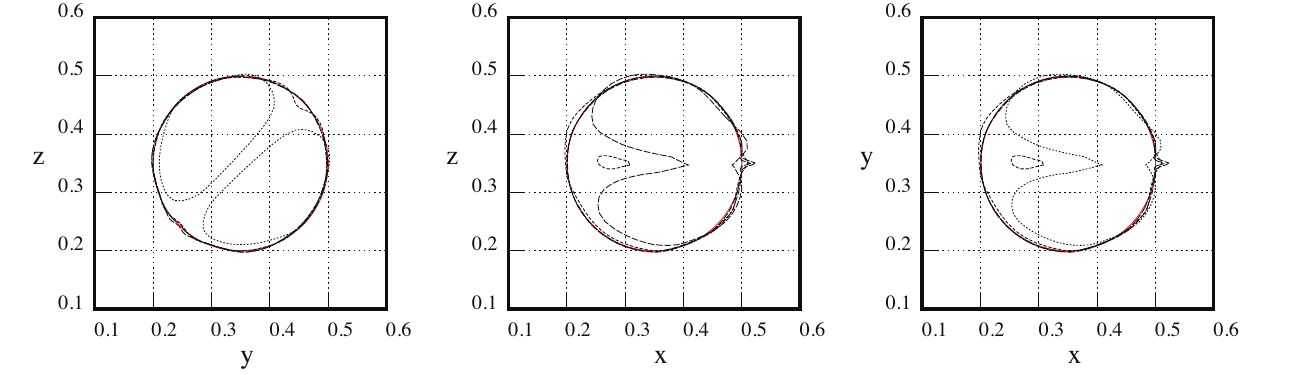}
  \caption{Contours for $\theta=0.5$ in $x=0.35$, $y=0.35$ and $z=0.35$ planes after $3.0$ time-units for meshes of $72^3$, $144^3$, $288^3$, $432^3$ \Changes[]{and $576^3$} cells.}
  \label{fig:contours}
\end{figure}

\begin{figure}[h]
  \centering
  \includegraphics[width=\textwidth]{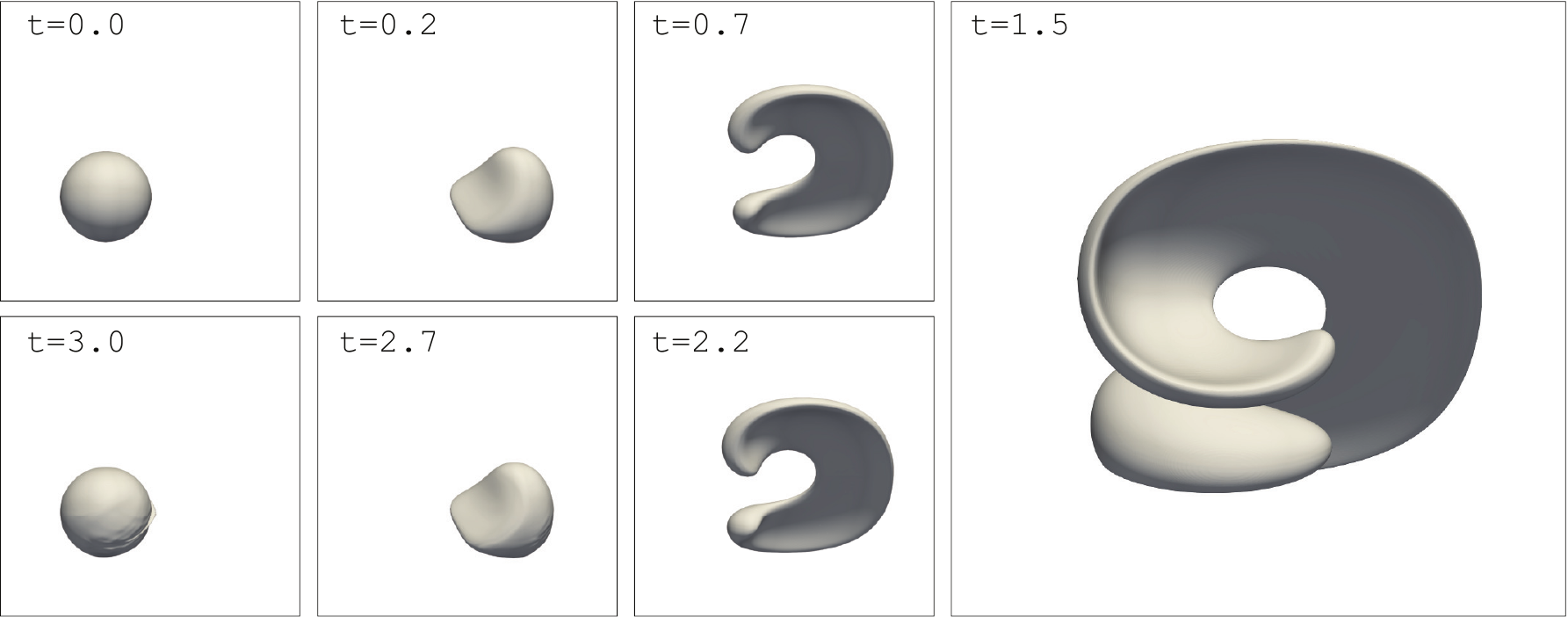}
  \caption{Time evolution of the $\theta=0.5$ contour for $t = 0$, $0.25$, $0.75$, $1.5$, $2.25$, $2.75$ and $3.0$ time-units for mesh of $576^3$ cells.}
  \label{fig:renders}
\end{figure}

Next, the application of this technique is applied to a canonical benchmark. In particular, the deformation (advection) of a sharp profile, which has been tested on \threed~hexahedral meshes of $72^3$, $144^3$, $288^3$, $432^3$ and $576^3$ cells following Algorithm \ref{alg:ADVECTION-ALG}, where $\mathsf{M} \in \Reals^{\nel{C} \times \nel{F}}$ is the divergence operator \cite{Trias2014} and $\mathsf{U}(\xuf) \in \Reals^{\nel{F} \times \nel{F}}$ is a diagonal matrix containing the velocities at faces. Recall we evaluate the products by diagonal matrices by means of \xkbin~calls.
\begin{algorithm}[h]
  \caption{Algorithm for the advection of a scalar field with a 1st order Euler method, using the algebraic implementation of a flux limiter.}
  \begin{algorithmic}[1]
    \Require $\xtcn$, $\xuf$, $dt$, $\DINC{C \to F}$, $\UUD$, $\OUD$, $\shift{C \to F}$, $\mathsf{M}$
    \Ensure $\xtcnn$
    \State $\mathrlap{\dfc}\hphantom{\xtcnn} \gets \SIGN \DINC{C \to F} \xtcn$
    \State $\mathrlap{\xrf}\hphantom{\xtcnn} \gets (\SIGN \UUD + \OUD)\xtcn/\dfc$
    \State $\mathrlap{\xtf}\hphantom{\xtcnn} \gets \shift{C \to F}\xtcn + \FLIM \dfc$
    \State $\mathrlap{\xtcnn}\hphantom{\xtcnn} \gets \xtcn + dt \mathsf{M} \mathsf{U}(\xuf) \xtf$
  \end{algorithmic}
  \label{alg:ADVECTION-ALG}
\end{algorithm}
The sharp profile is initialized in a physical domain of $[0,1] \times [0,1] \times [0,1]$ as a sphere of radius $r=0.15$, located at $(0.35, 0.35, 0.35)$ and subject to a divergence-free velocity field:
\begin{align}
  u &= 2sin^2(\pi x)sin(2\pi y)sin(2\pi z) cos(\pi t/T),\\
  v &= -sin(2\pi x)sin^2(\pi y)sin(2\pi z) cos(\pi t/T),\\
  w &= -sin(2\pi x)sin(2\pi y)sin^2(\pi z) cos(\pi t/T),
  \label{eqn:rhodonea}
\end{align}
during $3.0$ time-units, $T$ \cite{Yang2021}.
The results of the profile on meshes of $72^3$, $144^3$, $288^3$, $432^3$ \Changes[]{and $576^3$} cells are shown in Figure \ref{fig:contours} for the slices in $x=0.35$, $y=0.35$ and $z=0.35$ planes. \Changes[The computations have been performed on up to 64 nodes (3,072 cores) of the CPU-based MareNostrum 4 supercomputer at the Barcelona Supercomputing Center. Its nodes with two Intel Xeon 8160 CPUs (24 cores, 2.1 GHz, 6 DDR4-2666 memory channels, 128 GB/s memory bandwidth, 33 MB L3 cache) are interconnected through the Intel Omni-Path network (12.5 GB/s). The \xhpcs~\cite{Alvarez2018}, our framework designed for the efficient implementation of algebraic algorithms on hybrid supercomputers, achieves a sustained performance of up to 1.6 \xtflops, which corresponds to nearly 80\% of the theoretically achievable performance.]{}
The \threed~temporal evolution of the sphere on a mesh of $576^3$ \Changes[]{cells} is shown in Figure \ref{fig:renders}. As in \cite{Yang2021}, the resulting shapes after the deformation are satisfactory, and mass is exactly conserved. \Changes[In this case, the simulation has been carried out on 27 nodes of the Lomonosov-2 supercomputer at Lomonosov Moscow State University. Its hybrid nodes are equipped with one Intel Xeon E5-2697 v3 CPU (14 cores, 2.6 GHz, 4 DDR4-2133 memory channels, 68 GB/s memory bandwidth, 35 MB L3 cache) and one NVIDIA Tesla K40M GPU (12 GB of GDDR5 memory, 288 GB/s, PCIe 3.0 x16 -- 16GB/s), and are interconnected via InfiniBand FDR network (7 GB/s). Our code achieves a sustained performance of 0.9 \xtflops and 98.5\% of heterogeneous performance (\ie~the sum of CPU-only and GPU-only performances), corresponding to nearly 75\% of the maximum achievable performance.]{}
\subsection{\Changes[]{Performance analysis}}
\label{sec:Performance}
\begin{figure}[h]
  \centering
  \includegraphics[width=\textwidth]{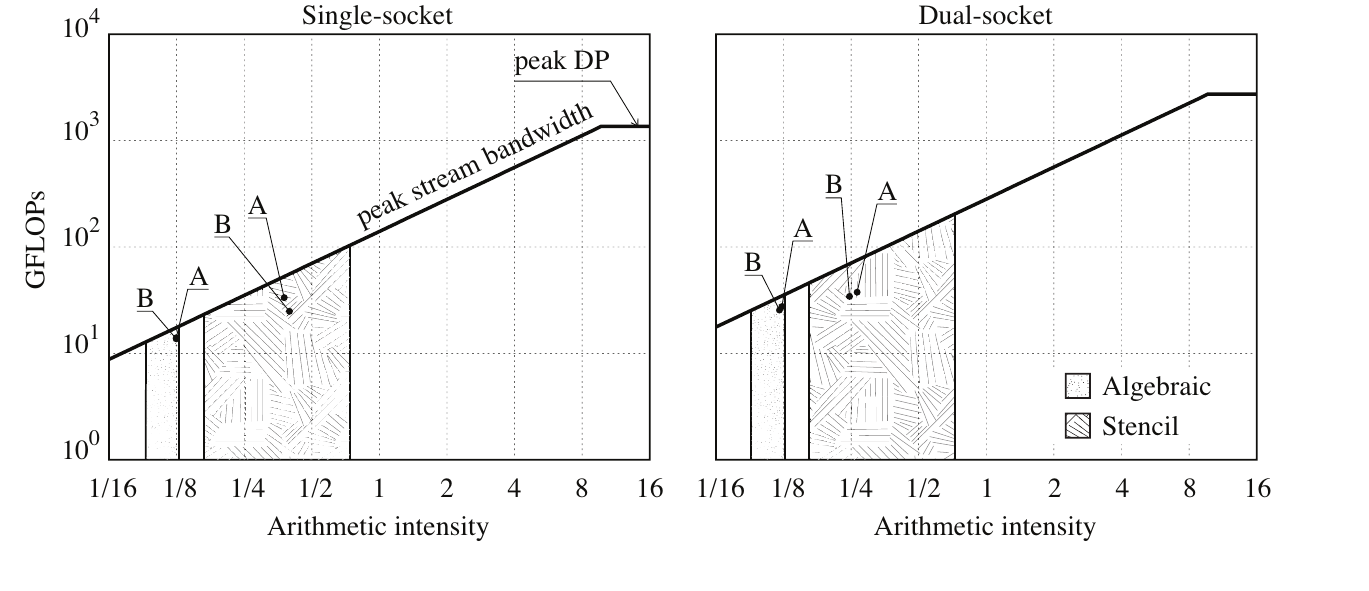}
    \caption{Roofline models representing the ranges of achievable performance for the stencil- and algebra-based implementations of the flux limiter on \protect\threed~unstructured meshes. Results on Intel Xeon Gold 6230 are shown for meshes of $144^3$ (A) and $288^3$ (B).}
  \label{fig:ROOFLINE}
\end{figure}

\Changes[]{The benchmark described in Section \ref{sec:Vortex} has been deployed on the \xhpcs~\cite{Alvarez2018} framework, designed for the efficient implementation of algebraic algorithms on hybrid supercomputers. To demostrate its portability, the simulations have been run on different supercomputers. Before going into details, we define the \emph{theoretical achievable performance} of a particular kernel, $\pi_k$, as:}
\begin{equation*}
 \pi_k = min(\pi, AI_k\beta),
\end{equation*}
\Changes[]{where $\pi$ is the peak performance of the computing device in double-precision, $\beta$ is the peak memory bandwidth and $AI_k$ is the maximum \xai~of the kernel, taking the full-hit scenario as described in Section \ref{sec:ComparisonWithStencil}. Then, we define the \emph{performance} and \emph{memory efficiency} as the ratio of measured performance to $\pi_k$ and measured memory traffic to full-hit, respectively.}
\Changes[]{The simulations on meshes of $72^3$--$432^3$ cells have been executed on up to 64 nodes (3,072 cores) of the CPU-based MareNostrum 4 supercomputer at the Barcelona Supercomputing Center. Its nodes are equipped with two Intel Xeon 8160 CPUs (24 cores, 2.1 GHz, 6 DDR4-2666 memory channels, 128 GB/s memory bandwidth, 33 MB L3 cache), interconnected through the Intel Omni-Path network (12.5 GB/s). The application achieved a sustained performance of up to 1.6 \xtflops, corresponding to nearly 80\% of performance efficiency.}
\Changes[]{The simulation on a mesh of $576^3$ cells has been executed on 27 nodes of the Lomonosov-2 hybrid supercomputer at Lomonosov Moscow State University. Its hybrid nodes are equipped with one Intel Xeon E5-2697 v3 CPU (14 cores, 2.6 GHz, 4 DDR4-2133 memory channels, 68 GB/s memory bandwidth, 35 MB L3 cache) and one NVIDIA Tesla K40M GPU (12 GB of GDDR5 memory, 288 GB/s, PCIe 3.0 x16 -- 16GB/s), interconnected via InfiniBand FDR network (7 GB/s). The application achieved a sustained performance of 0.9 \xtflops, corresponding to nearly 75\% of performance efficiency and 98.5\% of the heterogeneous efficiency (\ie~the ratio of the heterogeneous performance to the sum of CPU-only and GPU-only performances).}
\Changes[]{Finally, we complete the performance analysis, and the theoretical comparison in Section \ref{sec:ComparisonWithStencil}, by comparing the actual performance of the algebraic and stencil approaches. Following Algorithm~\ref{alg:FLUX-LIMITER-UNS}, a stencil kernel has also been implemented in \xhpcs. The tests have been carried out on the CPU-based JFF cluster at the Heat and Mass Transfer Technological Center. Its nodes are equipped with two Intel Xeon Gold 6230 CPUs (20 cores, 2.1 GHz, 6 DDR4-2933 memory channels, 140 GB/s memory bandwidth, 27.5 MB L3 cache). Considering that the \xdm~parallelization is equivalent in both approaches since the data exchanges are the same, only single-node comparisons have been conducted.}
\Changes[]{The results are shown in a roofline plot in Figure \ref{fig:ROOFLINE} for both single- and dual-socket executions using a \xnuma-aware, \xsm~parallelization on meshes of $144^3$ (A) and $288^3$ (B) cells. We have discarded the smallest mesh of $72^3$ to ensure a memory-bounded behavior and the biggest meshes of $432^3$ and $576^3$ because they do not fit in a single node. In the plot, two vertical lines represent the minimum and maximum values of the \xai~for each kernel as estimated in Section \ref{sec:ComparisonWithStencil} and outlined in Table \ref{tab:FLUX-LIMITER-REQUIREMENTS}. Then, the roofline curve is calculated as follows:}
\begin{equation*}
\Pi(\pi, \beta) = min(\pi, AI\beta).
\end{equation*}
\Changes[]{In this particular test, $\beta$ and $\pi$ are 140 GB/s and 1344 \xgflops~per socket, respectively. The real number of memory accesses to main memory have been measured using a profiling tool to calculate the effective AI of each execution.}
\Changes[]{In single-socket execution, the stencil kernel performs nearly twice faster than the algebraic one (stencil: 33.02 (A) and 24.52 (B) \xgflops, algebraic: 13.89 (A) and 13.54 (B) \xgflops), even though its lower performance efficiency (stencil: 32\% and 24\%, algebraic: 78\% and 76\%). However, this gap is reduced to approximately $\times 1.35$ in the dual-socket case (stencil: 37.01 and 33.82 \xgflops, algebraic: 27.16 and 25.17 \xgflops). The algebraic kernels feature a regular unit-strided memory access everywhere except the input vector in \xspmv. In contrast, the stencil kernel leads to irregular accesses to $\DINC{F \to C}$, $\DINC{C \to F}$, $\xdf$, $\xtc$ and $\vec{\dvec{n}}$, resulting in higher cache miss rates and reducing the memory efficiency, especially in dual-socket configurations (stencil: 36\% and 33\%, algebraic: 96\% and 94\%). Thus, the actual performance gap is far from the $\times 5.75$ of the (worst) theoretical scenario.}
For further details of the implementation of our framework and a detailed performance and scalability analysis on different types of supercomputing facilities, the reader is referred to \'Alvarez-Farr\'e et al. \cite{Alvarez2021}.

%% file: 5-Discussion.tex
\section{Discussion}
\label{sec:Discussion}
\begin{figure}[h]
  \centering
  \includegraphics[width=0.5\textwidth]{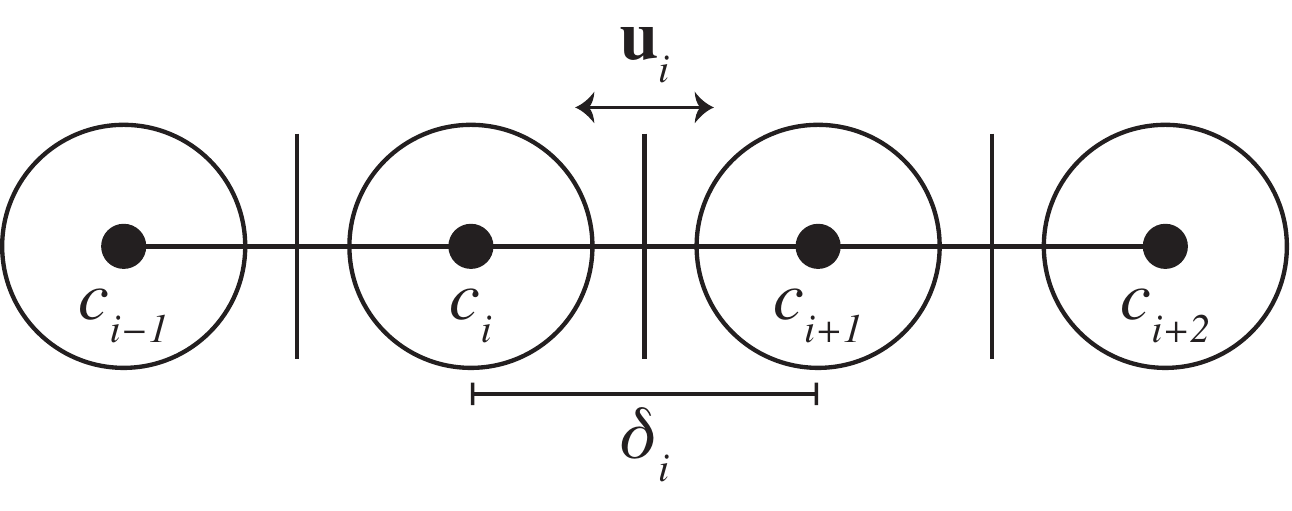}
  \caption{Equivalent stencils used in the computation of an algebraic flux limiter. In this case the adjacency matrices involve the operation with all neighboring nodes.}
  \label{fig:FLAlgebraic}
\end{figure}

\Changes[]{The advantages and disadvantages of the algebra- and stencil-based implementations of the flux limiter are discussed below.}
The algebraic formulation of flux limiters proposed in this work allows for fitting the calculation of such high-resolution schemes into algebra-based frameworks. Following this approach, the \xdns~and \xles~of turbulent multiphase flows, for instance, is reduced to a minimal set of algebraic subroutines, leading to fully portable and sustainable implementations. The challenges associated with the introduction of new architectures, \Changes[along with]{and} the ongoing hybridization of \xhpc~systems make \Changes[this]{these} two properties of the uttermost importance in the development of modern scientific computing codes. \Changes[]{Moreover, algebraic kernels are so widespread that optimized libraries are available for virtually all the existing architectures.}
From the results listed in Table \ref{tab:FLUX-LIMITER-REQUIREMENTS}, a significant \Changes[]{(theoretical)} overhead is revealed in the algebraic implementation. This overhead is mainly induced from Equation \ref{eqn:SimplifiedDeltaUpstream}, where we are enforced to calculate the contribution of all neighboring faces twice as described in Figure \ref{fig:FLAlgebraic}, according to the oriented and unoriented matrices, to cancel the downwind values. Note that $\OUD$ and $\UUD$ are the matrices with \Changes[]{the} largest number of non-zero elements. On the other hand, the algebraic implementation requires \Changes[a higher number of]{more} kernel calls which reduces the maximum \xai~of the algorithm. \Changes[However, although the differences in achievable performance are evident, various studies have shown that reaching the bandwidth limit with stencil kernels is hard due to the irregular access and complex operations \cite{Williams2009, Mostafazadeh2018}, in contrast with the simpler algebraic kernels \cite{Williams2009}.]{However, the memory access patterns in algebraic kernels is regular and unit-strided everywhere except the input vector in \xspmv.}
\Changes[]{Although the differences in theoretical achievable performance ($\pi_k$) are evident, reaching high performance and memory efficiencies with complex stencil kernels usually requires more complex optimizations. For instance, the authors in \cite{Williams2009} study different types of kernels, including \xspmv, and show that simpler kernels require fewer optimizations to reach higher efficiencies, even though their absolute performance is still lower. Also, the authors in \cite{Mostafazadeh2018} study the progressive performance improvement of a \xcfd~solver applying several optimizations such as kernel fusion, cache-blocking, vectorization, and \xnuma-aware memory initialization. Indeed, their fully-optimized solver reports absolute performances that are already higher than the $\pi_k$ of any equivalent algebraic counterpart. However, from their results, the performance gap is far from the (worst) theoretical predictions in Table \ref{tab:FLUX-LIMITER-REQUIREMENTS} because their stencil application achieves low performance efficiencies (around 10--40\%). In contrast, our measurements in Section \ref{sec:Performance} show that our algebraic kernels report a very stable 80\% efficiency. Furthermore, the test case in \cite{Mostafazadeh2018} on a mesh with 2 million grid points appears to be benefiting from cache reuse, especially in the Broadwell device with 56 MB L3 cache, hence the gap on larger grids should be even smaller.}
\Changes[]{Regarding the parallelization both implementations are very similar. The distributed-memory parallelization remains the same. In both cases, the overlapping schemes for hiding the communication overhead have reported solid results \cite{Alvarez2021, Gorobets2018}.}
Finally, it is noteworthy that the calculation of $\xtf$ represents a marginal part of a simulation, \Changes[]{and that the overhead of the algebraic implementation is not significant in other evaluations} such as the advection--diffusion of the variables. Indeed, the most time- and memory-consuming part in \xcfd~simulations is the solver of \xslae~which, in our implementation, not only does not suffer but benefits: our fields are vectors suitable for the \xslae~throughout the entire simulation. Neither copy, transfer, nor manipulation \Changes[are]{is} required for passing our vectors as input parameters to our solvers. \Changes[]{Therefore, the drawbacks of the algebraic approach of the flux limiter are diminished in an actual \xcfd~simulation.}

%% file: 6-Conclusions.tex
\section{Conclusions and Future Work}
\label{sec:Conclusions}
A flux limiter scheme has been formulated from an algebraic perspective resulting in a compact formulation that allows for \Changes[an]{} easy implementation on algebraic frameworks. The resulting implementation provides \Changes[with]{} accurate results and collapses to the traditional approach of Sweby \cite{Sweby1984} when a homogeneous, Cartesian grid is used.
Graph incidence matrices (both directed and undirected) are exploited to construct appropriate gradient ratios, while the face velocity sign determines the appropriate side to pick the upstream information. After the sign operation, the remaining operations are \Changes[all of them]{} either linear or local.
This approach presents several advantages, most remarkably \Changes[the reduction of]{reducing} the number of computing kernels that need to be ported when moving to new architectures. On the other hand, a theoretical comparison with respect to a classical stencil-based implementation reveals that the latter is cheaper regarding the memory traffic because it can make use of specialized kernels that require less intermediate results, or discriminate some operations with conditional statements, most remarkably when locating upwind values. \Changes[However, these specializations may harm the efficiency of memory transactions, which is very important in some parallel paradigms such as \xsp, and thus reduce the sustained performance which, in our framework, is near the theoretically achievable performance given by the memory bandwidth.]{However, the performance study shows that the performance gaps are much smaller than the worst theoretical scenario ($\times 1.35$ instead of $\times 5.75)$. Either way, the calculation of $\xtf$ represents a marginal part of an actual \xcfd~simulation and, therefore, the drawbacks of the algebraic approach in realistic simulations are diminished.}
Finally, the approach developed in this work can be improved to include the effect of the non-homogeneous distance across upstream faces or its surface in \Changes[the calculation of]{calculating} the upstream gradient.

%% file: acknowledgments.tex
\section{Acknowledgments}
\label{sec:Acknowledgments}
The work of N.~V. and X.~\'A.~F. has been supported by the Government of Catalonia, FI AGAUR predoctoral grants 2019FI\_B2\_000104 and  2019FI\_B2\_00076.
\Changes[A.~G has been funded by the Russian Science Foundation, project 19-11-00299.]{}
N.~V., X.~\'A.~F., J.~C., A.~O. and F.~X.~T. have been funded by the Spanish Research Agency, ANUMESOL project ENE2017-88697-R.
J.~C. has also been funded by Spanish Research Agency, GALIFLOW project ENE2015-70662-P.
The studies of this work have been carried out using the MareNostrum 4 supercomputer of the Barcelona Supercomputing Center, projects IM-2019-3-0026 and IM-2020-1-0006;
the TSUBAME3.0 supercomputer of the Global Scientific Information and Computing Center at Tokyo Institute of Technology;
the Lomonosov-2 supercomputer of the shared research facilities of \xhpc~computing resources at Lomonosov Moscow State University;
the K-60 hybrid cluster of the collective use center of the Keldysh Institute of Applied Mathematics;
the JFF cluster of the Heat and Mass Transfer Technological Center at Technical University of Catalonia.
The authors thankfully acknowledge these institutions for the compute time and technical support.